\documentclass[12pt]{article}
\pdfoutput=1

\usepackage{amsmath,amssymb,amsfonts}
\usepackage{psfrag}
\usepackage{color}
\definecolor{darkblue}{rgb}{0.1,0.1,.7}
\usepackage[colorlinks, linkcolor=darkblue, citecolor=darkblue, urlcolor=darkblue, linktocpage]{hyperref} 
\usepackage[square, comma, sort&compress,numbers]{natbib}
\usepackage[]{amsmath}
\usepackage[]{graphicx}
\usepackage[]{latexsym}
\usepackage{geometry}
\newcommand\SDPB{\texttt{SDPB}}
\usepackage{amscd}
\usepackage[all,cmtip]{xy}
\usepackage{mathrsfs}
\usepackage[margin=10pt,font=small,labelfont=bf]{caption}
\usepackage{dsdshorthand}
\usepackage{simplewick}
\usepackage{bbold}
\usepackage{changepage}
\usepackage{slashed}
\usepackage{pstool}
\usepackage{setspace}
\numberwithin{equation}{section}

\renewcommand{\be}{\begin{eqnarray}}
\renewcommand{\ee}{\end{eqnarray}}
\newcommand{\bea}{\begin{eqnarray}}
\newcommand{\eea}{\end{eqnarray}}

\def\beq{\begin{equation}} 
\def\eeq{\end{equation}} 
\def\<{\langle}
\def\>{\rangle}

\def\nn{\nonumber} 
 
\def\cO {{\cal O}}

\def\cN {{\cal N}}

\newcommand\numax{\kappa}

\newcommand{\es}[2] {\begin{equation} \label{#1} \begin{split} #2 \end{split} \end{equation}}

\begin{document}

\vspace*{-.6in} \thispagestyle{empty}
\begin{flushright}
PUPT-2524
\end{flushright}
\vspace{.2in} {\Large
\begin{center}
{\bf Bootstrapping 3D Fermions with Global Symmetries\\\vspace{.1in}}
\end{center}
}
\vspace{.2in}
\begin{center}
{\bf 
Luca Iliesiu$^{a}$,
Filip Kos$^{b,c}$, 
David Poland$^{d}$,\\
Silviu S.~Pufu$^{a}$,
David Simmons-Duffin$^{e,f}$
} 
\\
\vspace{.2in} 
$^a$ {\it  Joseph Henry Laboratories, Princeton University, Princeton, NJ 08544, USA}\\
$^b$ {\it  Berkeley Center for Theoretical Physics, Department of Physics,\\ University of California, Berkeley, CA 94720, USA}\\
$^c$ {\it  Theoretical Physics Group, Lawrence Berkeley National Laboratory, CA 94720, USA}\\
$^d$ {\it  Department of Physics, Yale University, New Haven, CT 06520, USA}\\
$^e$ {\it School of Natural Sciences, Institute for Advanced Study, Princeton, NJ 08540, USA}\\
$^f$ {\it Walter Burke Institute for Theoretical Physics, Caltech, Pasadena, CA 91125, USA}
\end{center}

\vspace{.2in}

\begin{abstract}
We study the conformal bootstrap for 4-point functions of fermions $\<\psi_i \psi_j \psi_k \psi_{\ell}\>$ in parity-preserving 3d CFTs, where $\psi_i$ transforms as a vector under an $O(N)$ global symmetry. We compute bounds on scaling dimensions and central charges, finding features in our bounds that appear to coincide with the $O(N)$ symmetric Gross-Neveu-Yukawa fixed points. Our computations are in perfect agreement with the $1/N$ expansion at large $N$ and allow us to make nontrivial predictions at small $N$. For values of $N$ for which the Gross-Neveu-Yukawa universality classes are relevant to condensed-matter systems, we compare our results to previous analytic and numerical results. 
\end{abstract}

\newpage

\begin{spacing}{0.8}
\tableofcontents
\end{spacing}

\newpage

\section{Introduction}
\label{sec:intro}

In recent years the conformal bootstrap~\cite{Polyakov:1974gs,Ferrara:1973yt,Mack:1975jr} has had striking success in constraining 3D conformal field theories (CFTs) of relevance to critical phenomena in statistical and condensed matter systems. These studies focused on systems of bootstrap equations for 4-point functions of scalar operators, leading to high precision determinations of critical exponents in models such as the 3D Ising~\cite{ElShowk:2012ht,El-Showk:2014dwa,Kos:2014bka,Simmons-Duffin:2015qma,Kos:2016ysd,Simmons-Duffin:2016wlq} and $O(N)$ vector models~\cite{Kos:2013tga,Kos:2015mba,Kos:2016ysd}. However, there are many CFTs of experimental interest containing fermionic excitations, which are unlikely to be isolated via the bootstrap by only considering scalar correlators. Such theories include the 3D Gross-Neveu-Yukawa (GNY) theories with $N$ Majorana fermions~\cite{Gross:1974jv, Iliesiu:2015qra, Gracey:1992cp,Derkachov:1993uw,Gracey:1993kc,Moshe:2003xn, Petkou:1996np}, the 3D Hubbard model on a honeycomb lattice~\cite{toldin2015fermionic}, models of graphene~\cite{herbut2006interactions, herbut2009relativistic} or of d-wave superconductors~\cite{vojta2000quantum, vojta2003quantum}, $\cN=1$~\cite{Iliesiu:2015qra} or $\cN=2$~\cite{bobev2015bootstrapping, chester2015accidental, Chester:2015lej} supersymmetric extensions of the Ising model, and more. 

In~\cite{Iliesiu:2015qra} we initiated a study of the conformal bootstrap applied to the 4-point function $\<\psi\psi\psi\psi\>$  of a single Majorana fermion $\psi$ of scaling dimension $\Delta_\psi$ that belongs to a 3D CFT with parity symmetry. We found that by varying the gap between the scaling dimensions $\Delta_\sigma$ and $\Delta_{\sigma'}$ of the first ($\sigma$) and the second ($\sigma'$) parity-odd scalar appearing in the $\psi \times \psi$ OPE, the resulting allowed region in the $\{\Delta_\psi, \Delta_\sigma\}$ plane showed features (kinks) that tracked the GNY fixed points at large $N$.  However, the precise value of $N$ corresponding to each of these kinks could be inferred only approximately from the value of $\Delta_{\sigma'}$.  From tracking the kinks as we varied $\Delta_{\sigma'}$ one can thus extract functional relations $\Delta_\sigma (\Delta_{\sigma'})$ and $\Delta_\psi(\Delta_{\sigma'})$, which we expect should hold, conjecturally, even for the GNY theories away from the large $N$ limit.  At some value of $\Delta_{\sigma'}$, the family of kinks intersects the supersymmetry line $\Delta_\psi = \Delta_{\sigma} + 1/2$, and we suspect that it is this value of $\Delta_{\sigma'}$ corresponds to the $N=1$ theory, which is expected to have enhanced ${\cal N} = 1$ supersymmetry \cite{Grover:2013rc,fei2016yukawa}. 

In order to study the GNY theories more precisely, we should make use of the information that these theories have $O(N)$ global symmetry.  In the present work, we will therefore generalize the study of~\cite{Iliesiu:2015qra} to 4-point functions of Majorana fermions $\<\psi_i \psi_j \psi_k \psi_{\ell}\>$, where $\psi_i$ transforms in the vector representation of an $O(N)$ global symmetry. In particular, we will use semidefinite programming methods to compute universal bounds on the leading $O(N)$ singlets and symmetric tensors appearing in the $\psi_i \times \psi_j$ OPE\@. As we will see, we observe a sequence of kinks in the space of allowed theories which match precisely with the GNY fixed points at large values of $N$. At small integer values of $N$, we use the locations of these kinks to make predictions for the critical exponents of these models. We additionally compute universal bounds for the stress-energy tensor and current central charges as a function of $N$, again finding features that coincide with the GNY models. 

When making no assumptions about the spectrum, the bounds on the scaling dimensions of $O(N)$ singlet operators do not make contact with the GNY models. (They do show, however, some evidence for the existence of new ``dead-end'' CFTs, as we will discuss.)  On the other hand, when we do make some plausible assumptions about the gap above the first parity odd singlet, we do find a set of kinks that correspond to expected scaling dimensions in the GNY models.  Furthermore, when imposing such a gap for small values of $N$, a second set of kinks is visible. This is reminiscent of structure found in $\epsilon$-expansion studies, e.g.~recently discussed in~\cite{fei2016yukawa}, where there is an additional fixed point besides the GNY model which for small values of $N$ has a possibility of becoming a unitary CFT\@. They are also similar to a second kink observed in~\cite{Iliesiu:2015qra}, appearing close to the feature conjectured to coincide with the $\cN=1$ supersymmetric Ising model. Our bootstrap results may support the existence of this second set of CFTs (which we refer to as GNY$^*$), though the story is currently unclear. The possible existence of GNY$^*$ theories is worth further study.

This paper is organized as follows. In Section~\ref{sec:GNY} we review the Gross-Neveu-Yukawa Model. In Section~\ref{sec:crossing} we discuss the crossing relations applicable to fermionic correlators with $O(N)$ symmetry and present the theoretical set-up for bootstrap applications. In Section~\ref{sec:GNY_spectrum} we specifically focus on the Gross-Neveu-Yukawa Model by either studying universal bounds on scaling dimensions or by placing bounds after imposing further gaps in the spectrum. In the latter case, we comment on a second set of kinks which solely appears for small values of $N$.  Next, in Section~\ref{sec:generic-bounds} we study an expanded set of universal bounds on scaling dimensions of operators and in Section~\ref{sec:ccbounds} we study universal bounds on central charges. Finally, in Section~\ref{sec:discussion} we discuss future directions.

\section{Review of Gross-Neveu-Yukawa Model}
\label{sec:GNY}

While most of the bootstrap bounds presented in this paper will have universal implications for the space of all CFTs, as mentioned in the Introduction, our main focus will be to study the CFT data of the Gross-Neveu model at criticality~\cite{Gross:1974jv}. The Gross-Neveu-Yukawa description contains $N$ Majorana fermions $\psi_i$ and a parity-odd scalar field $\phi$, with Lagrangian
   \be
    {\cal L} = -\frac 12 \sum_{i = 1}^N \bar \psi_i (\slashed{\partial} + g \phi ) \psi_i - \frac 12 \partial^\mu \phi\partial_\mu \phi  - \frac 12 m^2 \phi^2 - \lambda \phi^4 \,.
     \label{GNYLag}
   \ee
Here $g$ and $\lambda$ are coupling constants. The theory has an $O(N)$ global symmetry with $\psi_i$ transforming in the fundamental representation ($i = 1, \ldots, N$) in addition to a parity symmetry. For even values of $N$, this theory can be studied perturbatively in $d = 4 - \epsilon$ dimensions.  It has a critical point that can be achieved by appropriately tuning the scalar mass $m^2$, while parity symmetry forbids a fermionic mass term. This critical point, thought to survive down to $d=3$, has been extensively studied using perturbative, analytic and numerical methods.  Specifically, the model has been studied in the $1/N$ expansion~\cite{Gracey:1992cp,Derkachov:1993uw,Gracey:1993kc, Petkou:1996np, Moshe:2003xn, fei2016yukawa}, and in the $\epsilon$-expansion (see e.g.~\cite{ gracey1990three, rosenstein1993critical, zerf2016superconducting, Gracey:2016mio, fei2016yukawa, 2017arXiv170308801M} and references therein). The models have also been recently studied  from the perspective of weakly-broken higher spin symmetry in~\cite{Giombi:2017rhm}. The critical exponents at this fixed point have also been estimated through non-perturbative RG methods \cite{hofling2002phase, knorr2016ising, janssen2014antiferromagnetic} and numerically, using Monte Carlo simulations \cite{karkkainen1994critical, wang2014fermionic, li2015fermion, chandrasekharan2013quantum, hesselmann2016thermal}. 

Beyond serving as one of the most basic models for scalar-fermion interactions, the GNY-model and its variations frequently appear as universality classes for quantum phase transitions in condensed-matter systems with emergent Lorentz symmetry. It has been employed as a model for describing phase transitions in graphene~\cite{herbut2006interactions, herbut2009relativistic, 2017arXiv170308801M}, the Hubbard model on the honeycomb and $\pi$-flux lattice~\cite{toldin2015fermionic}, models of time-reversal symmetry-breaking in d-wave superconductors~\cite{vojta2000quantum, vojta2003quantum}, models of 3-dimensional gapless semiconductors~\cite{Moon:2012rx,Herbut:2014lfa} and models that exhibit emergent supersymmetry on the boundary of topological superconductors~\cite{Grover:2013rc}. Thus, we are optimistic that our results for this model can find future applications in a number of different experimentally-interesting systems. Specifically, for those interested in the application of bootstrap results to such systems, in Table~\ref{tab:exp-relev} we will list the critical exponents for the universality classes relevant to the metal-insulator transition for spinless fermions on the honeycomb lattice ($N=4$) and for the semi-metallic to insulator transition  in graphene ($N=8$). For convenience, Table~\ref{tab:exp-relev} also lists the values for the critical exponents obtained through several other methods. 

In the context of this work, we consider a four-point function of fermionic operators $\psi_i$. We can distinguish the operators appearing in the $\psi_i \times \psi_j$ OPE by their $O(N)$ representation: they can be in either the singlet representation ($S$), the two-index symmetric traceless representation ($T$), or the two-index anti-symmetric representation ($A$).   The dimensions of these operators can be estimated at large $N$ or in the $4-\epsilon$ expansion---see Table~\ref{tab:grossNeveuDimensions} for such estimates for a few of the lowest dimension operators.  In Table~\ref{tab:grossNeveuDimensions} we also give the large-$N$ estimates for the central charges of the stress-energy tensor and of the conserved $O(N)$ current, whose normalization we define in Section~\ref{sec:ccbounds}. 

\begin{table}[!t]
\centering
 Perturbative estimates for lowest-lying scalars in the GNY model
\begin{tabular}{c c c c c}
\hline\hline
 & \hspace{0.3cm} $\Z_2$ \hspace{0.3cm} & \hspace{0.3cm} $O(N)$  \hspace{0.3cm} &   \hspace{0.3cm}$\De$(large $N$) &   \hspace{0.3cm}$\De$($\epsilon$-expansion)\\
 \hline 
\hspace{0.5cm} $\psi_i$ \hspace{0.5cm} & $+$ & $V$ & $1+\frac{4}{3\pi^2 N} + \dots$ & $\frac{3}2 - \frac{N+5}{2(N+6)} \epsilon + \dots$\\
$\f$ & $-$ & $S$ & $1-\frac{32}{3\pi^2 N} + \dots$&  $1-\frac{3}{N+6} \epsilon+ \dots$ \\
$\f^2$ & $+$ & $S$ & $2+\frac{32}{3\pi^2 N}+\dots$& $2+ \frac{\sqrt{N^2 +132 N + 36} - N-30}{6(N+6)}\epsilon + \dots$\\
$\f^3$ & $-$ & $S$ & $3+\frac{64}{\pi^2 N}+\dots$&$3+ \frac{\sqrt{N^2 +132 N + 36} - N-12}{2(N+6)}\epsilon + \dots$ \\
$\bar\psi_{(i}\psi_{j)}$ & $-$ & $T$ & $2+\frac{32}{3\pi^2 N}+\dots$& -\\
$J^\mu \phi^2 \partial_\mu \partial^2 \phi $ & $-$ & $A$ & $8+\dots$& -\\
\hline\hline
\end{tabular}\\\vspace{0.5cm}
 Perturbative estimates for central charges in the GNY model\\
\begin{tabular}{c c c c l}
\hline\hline
  & $\hspace{0.5cm}\Z_2$ \hspace{0.5cm} &\hspace{0.5cm} $O(N)$ \hspace{0.5cm}& \hspace{0.5cm} $\De$ \hspace{0.5cm} & \hspace{0.5cm} \text{Central charge (large $N$)}\\
 \hline
\hspace{0.5cm}$J^\mu$ \hspace{0.5cm}& $+$ & $A$ & $2$ & $C_J = 1 - 
\frac{64}{9\pi^2N} + \dots$\\
$T^{\mu \nu}$  & $+$ & $S$ & $3$& $C_T = N\left(1+ \frac{8}{9\pi^2 N}+\dots\right) $\\
\hline\hline
\end{tabular}
\caption{\textit{Top:}
Representations and one-loop dimensions of low-lying operators in the 3D GNY models.  $V$, $S$, $T$, $A$ denote the vector, singlet, rank-two traceless symmetric tensor, and rank-two antisymmetric tensor representations of $O(N)$, respectively.  At large $N$, the dimensions of $\psi_i$ was computed in~\cite{Gracey:1992cp,Derkachov:1993uw,Gracey:1993kc}, while the dimension of $\bar\psi_{(i}\psi_{j)}$ was computed in Appendix B of \cite{Iliesiu:2015qra}. We list no corrections for dimension of the lowest operator in the $O(N)$ anti-symmetric representation, as we could not find any discussion of it in the literature. The most recent $\epsilon$-expansion estimates for the scaling dimensions are available in \cite{2017arXiv170308801M}, where $O(\epsilon^3)$ corrections are available for the dimensions associated to $\psi_i$ and $\phi$. Prior work on the $\epsilon$-expansion for the GNY models can be found in \cite{gracey1990three, rosenstein1993critical, zerf2016superconducting, Gracey:2016mio, fei2016yukawa, 2017arXiv170308801M}.   \textit{Bottom:} Representations and one-loop values for the central charges for the $O(N)$ conserved current and for the stress-energy tensor at large $N$ as determined in~\cite{diab2016c_j}. }
\label{tab:grossNeveuDimensions}
\end{table}

As briefly mentioned in the Introduction, in the $4-\epsilon$ expansion, besides the critical GNY model one finds an additional fixed point (GNY$^*$) \cite{fei2016yukawa}. For large values of $N$, the additional fixed point has a negative $\lambda$ coupling in the Lagrangian (\ref{GNYLag}) and thus it is expected that GNY$^*$ is non-unitary at large $N$. In fact, when computing scaling dimensions at such a critical point, one finds that for large $N$, there are scaling dimensions in the theory that become negative, thus violating the unitarity bound. However, as $N$ is decreased, the scaling dimensions of such operators grow and eventually the unitarity bound may be satisfied. Thus, it is possible that at small values of $N$, the GNY$^*$  theories become unitary and could be detected using the conformal bootstrap. Generically, one expects that the scaling dimension of the scalar $\phi$ in the theory is lower in GNY$^*$ than in GNY, such that the operator $\phi^4$ is relevant in GNY$^*$, but irrelevant in GNY, and one could thus flow from GNY$^*$ in the UV to the GNY model in the IR\@. As we discuss below, the GNY$^*$ models may appear in bootstrap bounds on the dimension of low-lying parity-odd operators.

\section{Crossing and Bootstrap with $O(N)$ Symmetry}
\label{sec:crossing}

In this section we set up the conformal bootstrap constraints for 4-point functions of Majorana fermions, $\psi_i$, transforming in the fundamental vector of a global $O(N)$ symmetry of a parity-preserving 3d CFT\@. To keep our formulas compact, we follow the notation of~\cite{Iliesiu:2015qra} and contract the fermions with auxiliary commuting polarization variables $s_{\alpha}$. By explicitly imposing the  $O(N)$ global symmetry, we will thus generalize the analysis of~\cite{Iliesiu:2015qra}, where we focused on the case of four identical Majorana fermions.

Three 3-point functions between two fermions and a spin-$\ell$ operator $\cO_{\ell}$, 
\be
\<\psi_1(x_1,s_1) \psi_2(x_2,s_2) \cO_{\ell}(x_3,s_3)\> \propto \sum_a \lambda^a_{\cO} r_a \,,
\ee
have four possible tensor structures $r_a$ with independent coefficients $\lambda^a_{\cO}$, of which two ($a=1,2$) have even parity and two ($a=3,4$) have odd parity.\footnote{These structures are simple to count using the formalism of \cite{Kravchuk:2016qvl}.} In addition, we can work in a basis such that structures $a=1,2,3$ are anti-symmetric under the exchange $1 \leftrightarrow 2$ if $\ell$ is even and symmetric if $\ell$ is odd, while the structure $a=4$ is symmetric under $1 \leftrightarrow 2$ if $\ell$ is even and anti-symmetric if $\ell$ is odd.

Similarly, 4-point functions of fermions have in general eight tensor structures $t_I$. In the case that the fermions transform as vectors under an $O(N)$ global symmetry, the conformal block decomposition of the 4-point function can be organized in terms of representations of $O(N)$:
\begin{align}
&\left(  x_{12}^{2\De_{\psi}+1} x_{34}^{2\De_{\psi}+1} \right)\<\psi_i(x_1,s_1) \psi_j(x_2,s_2)\psi_k(x_3,s_3) \psi_{\ell}(x_4,s_4)\> = \sum_{I=1}^8 t_I \Bigg\{   \notag\\ 
&\delta_{ij} \delta_{kl} \! \left[
\sum_{\substack{\cO \in S^{+},\, \ell \text{ even} \\ a, b = 1, 2}} \lambda_{\cO}^a \lambda_{\cO}^b g^I_{ab}(u,v) + \sum_{\cO \in S^{-},\, \ell \text{ even} } (\lambda_{\cO}^3)^2 g^I_{33}(u,v) + \sum_{\cO \in S^{-},\, \ell \text{ odd} } (\lambda_{\cO}^4)^2  g^I_{44}(u,v)
\right] \notag\\
+&(\delta_{ik} \delta_{jl} \!+\! \delta_{il} \delta_{jk} \! -\! \frac{2}{N}\delta_{ij} \delta_{kl}   )  \mkern-8mu \left[
\sum_{\substack{\cO \in T^{+},\, \ell \text{ even} \\ a, b = 1, 2}} \mkern-18mu \lambda_{\cO}^a \lambda_{\cO}^b g^I_{ab}(u,v) + \mkern-18mu \sum_{\cO \in T^{-},\, \ell \text{ even} } \mkern-18mu (\lambda_{\cO}^3)^2 g^I_{33}(u,v) + \mkern-18mu \sum_{\cO \in T^{-},\, \ell \text{ odd} } \mkern-18mu (\lambda_{\cO}^4)^2  g^I_{44}(u,v)
\right] \notag\\
+&(\delta_{ik} \delta_{jl}  -\delta_{il} \delta_{jk} ) \! \left[
\sum_{\substack{\cO \in A^{+},\, \ell \text{ odd} \\a, b = 1, 2}} \mkern-18mu \lambda_{\cO}^a \lambda_{\cO}^b g^I_{ab}(u,v) + \mkern-18mu \sum_{\cO \in A^{-},\, \ell \text{ odd} }\mkern-18mu (\lambda_{\cO}^3)^2 g^I_{33}(u,v) + \mkern-18mu \sum_{\cO \in A^{-},\, \ell \text{ even} }\mkern-18mu  (\lambda_{\cO}^4)^2  g^I_{44}(u,v)
\right]\Bigg\} \,.\label{eq:crossing-eq-ugly}
\end{align}
Here $S$, $T$ and $A$ denote $O(N)$ singlets, two-index symmetric traceless tensors and two-index antisymmetric tensors. The superscript $\pm$ denotes whether the operators appearing are parity even or parity odd. So, for example, the first sum in the second line runs over $\{S^+, \ell \text{ even} \}$, which are parity even $O(N)$ singlet operators of even spin. Note that we work in a basis such that the structures $t_I$ are symmetric under exchange $1\leftrightarrow 3$ for $I=1,2,3,4$ and antisymmetric for $I=5,6,7,8$. Additionally, when external dimensions are equal, the $t_{3,4,8}$ contributions vanish. For detailed definitions of the tensor structures $r_a$ and $t_I$, see~\cite{Iliesiu:2015qra}.

The crossing equation under the exchange $1\leftrightarrow 3$ can then be written as:
\bea
\label{eq:first-crossing-eq}
\mathcal S_\text{A}^I + \left(1-\frac{2}{N} \right)\mathcal T_\text{A}^I - \mathcal A_\text{A}^I =0 \,,\\ \label{eq:second-crossing-eq}
\mathcal S_\text{B}^I + \left(-1-\frac{2}{N}\right)\mathcal T_\text{B}^I + \mathcal A_\text{B}^I =0 \,,\\ \label{eq:third-crossing-eq}
\mathcal T_\text{A}^I + \mathcal A_\text{A}^I =0 \,,
\eea
where
\be
\mathcal S_\text{A}^I  = \begin{cases} 
\sum \limits_{\substack{\cO \in S^{+},\, \ell \text{ even} \\a, b = 1, 2}} \lambda_{\cO}^a \lambda_{\cO}^b F_{ab,\De,\ell}^{+ I} +\mkern-8mu \sum \limits_{\cO \in S^{-},\, \ell \text{ even} } (\lambda_{\cO}^3)^2 F_{33,\De,\ell}^{+I} +\mkern-8mu \sum \limits_{\cO \in S^{-},\, \ell \text{ odd} } (\lambda_{\cO}^4)^2 F_{44,\De,\ell}^{+I} &\mbox{if } I=1,2 \\ 
\sum \limits_{\substack{\cO \in S^{+},\, \ell \text{ even} \\ a,b =1, 2}} \lambda_{\cO}^a \lambda_{\cO}^b F_{ab,\De,\ell}^{- I} + \mkern-8mu\sum \limits_{\cO \in S^{-},\, \ell \text{ even} } (\lambda_{\cO}^3)^2 F_{33,\De,\ell}^{-I} +\mkern-8mu \sum \limits_{\cO \in S^{-},\, \ell \text{ odd} } (\lambda_{\cO}^4)^2 F_{44,\De,\ell}^{-I} & \mbox{if } I=5,6,7,
\end{cases} 
\ee
and
\be
\mathcal S_\text{B}^I  = \begin{cases}
\sum \limits_{\substack{\cO \in S^{+},\, \ell \text{ even} \\a, b =1, 2}} \lambda_{\cO}^a \lambda_{\cO}^b F_{ab,\De,\ell}^{- I} + \mkern-8mu \sum \limits_{\cO \in S^{-},\, \ell \text{ even} } (\lambda_{\cO}^3)^2 F_{33,\De,\ell}^{-I} + \mkern-8mu \sum \limits_{\cO \in S^{-},\, \ell \text{ odd} } (\lambda_{\cO}^4)^2 F_{44,\De,\ell}^{-I}  &\mbox{if }  I=1,2 \\ 
 \sum \limits_{\substack{\cO \in S^{+},\, \ell \text{ even} \\a, b=1, 2 }} \lambda_{\cO}^a \lambda_{\cO}^b F_{ab,\De,\ell}^{+ I} +  \mkern-8mu \sum \limits_{\cO \in S^{-},\, \ell \text{ even} } (\lambda_{\cO}^3)^2 F_{33,\De,\ell}^{+I} +  \mkern-8mu \sum \limits_{\cO \in S^{-},\, \ell \text{ odd} } (\lambda_{\cO}^4)^2 F_{44,\De,\ell}^{+I} &\mbox{if } I=5,6,7.
\end{cases}
\ee
Similar definitions apply to $\mathcal T^{I}_{A/B}$ and $\mathcal A^{I}_{A/B}$, which sum the contributions of $O(N)$ symmetric tensors and antisymmetric tensors, following the pattern of quantum numbers appearing in (\ref{eq:crossing-eq-ugly}). The functions $F_{ab,\De,\ell}^{\pm I} $ are defined as
\be
F_{ab,\De,\ell}^{\pm I} \equiv v^{\De_{\psi}+\frac12} g_{ab,\De,\ell}^{I}(u,v) \pm u^{\De_{\psi}+\frac12} g_{ab,\De,\ell}^{I}(v,u)\,.
\ee

We can now exclude assumptions about the spectrum by applying a set of functionals $\vec \alpha_{I}$  to equations \eqref{eq:first-crossing-eq}--\eqref{eq:third-crossing-eq}, 
\be
0 = \sum_{I, R} \left[\sum_{\substack{\cO^+_R, \,\ell\in \ell^R\\a,b=1,2}} \lambda^a_{\cO^+_R} \lambda^b_{\cO^+_R} \vec \alpha_{I} \cdot \vec V^{I, R}_{ab,\De,\ell} + \sum_{\cO^-_R, \,\ell\in \ell^R} (\lambda^3_{\cO^-_R})^2 \vec \alpha_{I}\cdot\vec V^{I, R}_{33,\De,\ell} + \sum_{\cO^-_R, \,\ell\in \bar\ell^R} (\lambda^4_{\cO^-_R})^2 \vec \alpha_{I}\cdot\vec V^{I, R}_{44,\De,\ell} \right] \,, \notag\\
\label{eq:crossing-summarized}
\ee
where the sum is over all $O(N)$ representations $R = S,\, T, \text{ or } A$. Here the sets of possible spins allowed in each representation, $\ell^R$ and their complements $\bar \ell^R$, are given by $\ell^S = \ell^T = \bar \ell^A = \{\text{all even spins}\}$ and $\bar \ell ^S = \bar \ell ^T = \ell^A  =  \{\text{all odd spins}\}$.  The vectors $\vec V_{ij}^{I, R}(u, v)$, are obtained from Eq.~(\ref{eq:crossing-summarized}), 
\be
\vec V_{ij}^{I, S}= \left(\begin{array}{c}
\mathcal{S}_{A, ij, \Delta, \ell}^I\\
\mathcal{S}_{B, ij, \Delta, \ell}^I\\
0
\end{array}\right),
\,\,\,\, \vec V_{ij}^{I, T} = \left(\begin{array}{c}
\left(1-\frac{2}N\right)\mathcal{T}_{A, ij, \Delta, \ell}^I\\
\left(-1-\frac{2}N\right)\mathcal{T}_{B, ij, \Delta, \ell}^I\\
\mathcal{T}_{A, ij, \Delta, \ell}^I
\end{array}\right),\,\,\,\,  \vec V_{ij}^{I, A} = \left(\begin{array}{c}
-\mathcal{A}_{A, ij, \Delta, \ell}^I\\
\mathcal{A}_{B, ij, \Delta, \ell}^I\\
\mathcal{A}_{A, ij, \Delta, \ell}^I
\end{array}\right),\label{eq:def-of-Vs}
\ee
where $\mathcal{S}_{A/B, ij, \Delta, l}^{I}$, $\mathcal{T}_{A/B, ij, \Delta, l}^{I}$, and $\mathcal{A}_{A/B, ij, \Delta, l}^{I}$ correspond to the contribution of the conformal block $F_{ij, \Delta, l}^I$ to $\mathcal S_{A/B}^I$, $\mathcal T_{A/B}^I$, and $\mathcal A_{A/B}^I$, respectively. 

Concretely, we look for a vector of functionals $\vec\alpha_{I}$ that satisfies the inequalities:
\be
-\sum_{a,b=1,2} \lambda^a_{\mathbb{1}} \lambda^b_{\mathbb{1}} \vec \alpha_{I}\cdot \vec V^{I, S}_{ab,0,0}&>& 0 \,,  \nonumber\\
\vec \alpha_{I} \cdot \vec V^{I, R}_{ab,\De,\ell} &\succeq& 0,\qquad \text{for all } R \text{ with } \ell \in  \ell^R \,, \nonumber\\
\vec \alpha_{I}\cdot \vec V^{I, R}_{33,\De,\ell} &\geq& 0,\qquad \text{for all } R \text{ with } \ell \in  \ell^R \,, \nonumber\\
\vec \alpha_{I}\cdot \vec V^{I, R}_{44,\De,\ell} &\geq& 0,\qquad \text{for all } R \text{ with } \ell \in \bar \ell^R \,,\label{eq:properties}
\ee
where we apply the functionals to blocks corresponding to all representations $R = S,\, T, \text{ or } A$. The inequalities (\ref{eq:properties}) should hold for all values of the dimensions $\De$ of operators present in the spectrum. We always assume the theory is a unitary CFT, which places a lower bound on the dimensions $\De \ge \De_{\text{min},\ell}$. Beyond unitarity, we can impose further gaps in the scaling dimensions of operators in each $O(N)$ representation and parity sector, leading to upper bounds on the scaling dimension of the lowest-lying operator in each sector if a suitable functional can be found.  Specifically, if such a vector of functionals $\vec\alpha_{I}$ exists when imposing a gap, then the crossing Eqs.~\eqref{eq:first-crossing-eq}--\eqref{eq:third-crossing-eq} cannot be satisfied and a CFT with such scaling dimensions cannot exist. Since in our conventions, all $\l_{\cO}^a$ are pure imaginary, we have a negative sign in the first line above. The OPE coefficients of the unit operator are given by $\l_{\mathbb{1}}^a = i\de^a_1$.  

Furthermore, in order to bound the OPE coefficient of a specific operator $\cO$ with dimension $\Delta_\cO$, spin $\ell_\cO$, and $O(N)$ representation $R_\cO$, we follow similar reasoning as above. We now search for a functional $\alpha$ such that:
\be
\label{eq:constraintsonfunctionalforcentralcharge}
-\sum_{(i,j)\in \mathcal I_\cO} \lambda^i_{\cO_\text{can}} \lambda^j_{\cO_\text{can}} \vec\alpha_{I}\cdot \vec V^{I, R_\cO}_{ij,\Delta_\cO, \ell_\cO } &=& 1 \,, \nonumber\\
\vec \alpha_{I}\cdot \vec V^{I, R}_{ab,\De,\ell} &\succeq& 0,\qquad \text{for all } R \text{ with } \ell \in  \ell^R \,,\nonumber\\
\vec \alpha_{I}\cdot \vec V^{I, R}_{33,\De,\ell} &\geq& 0,\qquad \text{for all } R \text{ with } \ell \in  \ell^R \,, \nonumber\\
\vec \alpha_{I}\cdot \vec V^{I, R}_{44,\De,\ell} &\geq& 0,\qquad \text{for all } R \text{ with } \ell \in \bar \ell^R \,,
\ee
where $\mathcal I_\cO$ gives the set of structures to which the operator $\cO$ contributes and the $\lambda^{i}_{\cO_{\text{can}}}$ are determined under some canonical normalization for the operator $\cO$ and are related to the OPE coefficients appearing in Eq.~\eqref{eq:crossing-summarized} by $\lambda^{i}_{\cO_{\text{can}}}/\lambda^{i}_{\cO} = \lambda^{j}_{\cO_{\text{can}}}/\lambda^{j}_{\cO}$ for all $(i, j) \in \mathcal I_\cO$. For instance, if we want to bound the OPE coefficient of the stress-energy tensor, as we do in Section~\ref{sec:ccbounds}, we will have $R_\cO =S$, and since the operator is parity-even, we have $\mathcal I_\cO = \{1,2\} \times \{1, 2\}$. In this case, the canonically normalized OPE coefficients will be determined by the Ward identity for the stress-energy tensor. 

 Eqs.~(\ref{eq:crossing-summarized}) and (\ref{eq:constraintsonfunctionalforcentralcharge}) then imply the inequality:
\be
\left(\frac{\lambda^i_{\cO}}{\lambda^i_{\cO_{\text{can}} }}\right)^2\le - \vec \alpha_{I} \cdot \vec V_{11,0,0}^{I}\,,
\ee
where the expression on the RHS corresponds to the identity operator contribution. Finding a functional $\vec \alpha_I$ obeying \eqref{eq:constraintsonfunctionalforcentralcharge} places an upper bound on $(\lambda_{\cO}^i/\lambda^i_{\cO_{\text{can}}})^2$. To make the bound as strong as possible, we search for an $\vec \alpha_I$ satisfying the relations \eqref{eq:constraintsonfunctionalforcentralcharge} that minimizes  $- \vec \alpha_I\cdot \vec V_{11,0,0}^{I}$.

We search for functionals satisfying either the (\ref{eq:properties}) or (\ref{eq:constraintsonfunctionalforcentralcharge}) constraints by approximating the search as a semidefinite program and implementing it in the solver \SDPB\ \cite{Simmons-Duffin:2015qma}, following the steps described in \cite{Iliesiu:2015qra}. A more detailed description of our  \SDPB\ implementation is presented in Appendix \ref{app:sdpb},  while the resulting constraints on the space of CFTs are presented below.

\section{Bootstrapping GNY Models from Scaling Dimension Bounds}
\label{sec:GNY_spectrum}
    \begin{figure}[t!]
    \centering
    Kinks corresponding to the GNY model when  bounding  $\Delta_{\sigma_T}$
    
\includegraphics[width=0.85\textwidth]{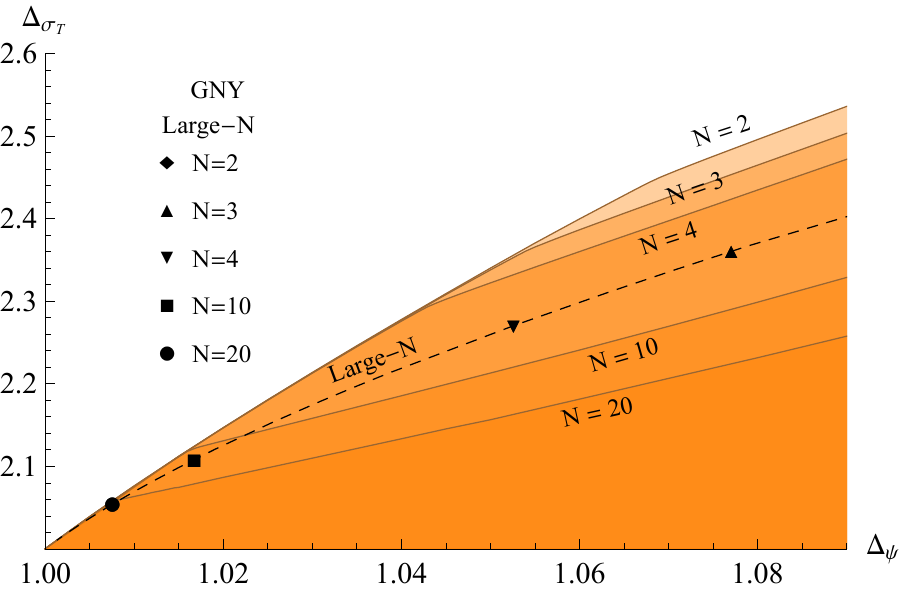}
    \caption{\label{fig:univ-bounds-sigmaT-zoomIn} Upper bounds on the scaling dimension of the lowest-lying parity-odd $O(N)$ symmetric-traceless tensor, $\Delta_{\sigma_T}$, for a unitary CFT containing fermions with scaling dimension $\Delta_\psi$.  We focus on the cases $N=2,\, 3,\,4, \, 10$ and $20$. The black symbols show the estimated values of the scaling dimensions $\Delta_{\psi}$ and $\Delta_{\sigma_T}$ obtained from the large-$N$ expansion up to $O(1/N^4)$ corrections \cite{fei2016yukawa}.  For $N=10,\, 20$ we note strong agreement.   This figure is a zoomed in version of Figure~\ref{fig:univ-bounds-sigmaT}. 
  }
  \end{figure}

    \begin{figure}[t!]
    \centering
    Bounds on scaling dimension of lowest $O(N)$ singlet parity-odd operator $\Delta_{\sigma}$, when imposing a gap above it in this sector up to the scaling dimension $\Delta_{\sigma'} > 3$ 
    
\includegraphics[width=0.99\textwidth]{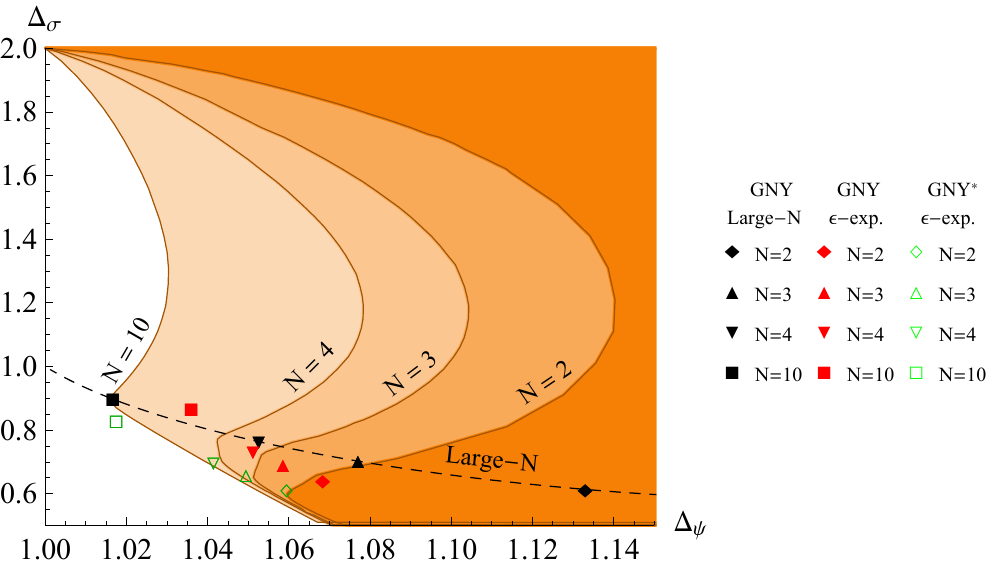}
    \caption{\label{fig:boundsWhenImposingGaps} Bound on the scaling dimension of the lowest-lying parity-odd $O(N)$ singlet, $\Delta_{\sigma}$, for a unitary CFT containing fermions with scaling dimension $\Delta_\psi$, when imposing a gap above this operator up to $\Delta_{\sigma'} > 3$.  Once again, we focus on the cases $N=2,\, 3,\,4,$ and $10$.  We notice that when imposing such a gap, we observe features that are close to the values of $\Delta_{\psi}$ found for the GNY kinks in Figure~\ref{fig:univ-bounds-sigmaT-zoomIn}. Estimates from the large-$N$ expansion (black markers) for the scaling dimension of $\Delta_{\sigma}$ also agree well with the position of the kink for $N=10$, but is inaccurate at smaller values of $N$. The red markers are the three-loop $\epsilon$-expansion results for the dimensions of the GNY models after performing a Pade$_{[2,1]}$ approximation  (see Eqs.~(11)--(13) and Table II in \cite{2017arXiv170308801M}). They are reasonably close to the upper kinks in the bounds for small $N$. The lower kinks appear close to the three-loop $\epsilon$-expansion estimates for the GNY$^*$ models (green hollow markers), obtained after performing a Pade$_{[1,2]}$ approximation, following the methods in \cite{fei2016yukawa} and \cite{2017arXiv170308801M}. While these second kinks are close to the $\epsilon$-expansion estimates for small $N$, for $N=10$ the second kink does not exist at all.  }
  \end{figure}

In this section we start off by presenting several selected bounds on scaling dimensions of scalars appearing in the $\psi_i\times\psi_j$ OPE\@. Let us first   focus on the most interesting bounds in the context of GNY models and leave a more systematic presentation of general bounds for Section~\ref{sec:generic-bounds}. Due to the intensive computation required for each individual run, in this work we will present full bounds for a limited set of values of $N$ ($N = 2, 3, 4, 10, 20$), as well as some additional GNY results at $N=8$.

We label the lowest dimension $O(N)$-invariant scalars in $\psi_i\times\psi_j$ OPE by $\sigma$ (parity odd) and $\epsilon$ (parity even). As before, we use transcripts $T$ and $A$ for operators transforming as $O(N)$ symmetric traceless tensors and $O(N)$ antisymmetric tensors, respectively. We have seen in Section~\ref{sec:crossing} that $\sigma, \sigma_T, \sigma_A, \epsilon, \epsilon_T$ can all appear in $\psi_i\times\psi_j$ OPE, but $\epsilon_A$ cannot. Higher dimension scalars in a given representation are labelled by increasing number of primes (e.g., $\sigma ', \sigma '', \dots$).

In Figure~\ref{fig:univ-bounds-sigmaT-zoomIn} we show the most general upper bounds on the dimension of the symmetric tensor $\sigma_T$ as a function of $\Delta_\psi$ for different values of $N$. For a range of $\Delta_\psi$ near 1, the bounds overlap and then depart from the shared curve at a critical value of $\Delta_\psi$. In this sense, all of the bounds seem to have the feature of a ``kink,'' reminiscent of the Ising model kink observed in \cite{ElShowk:2012ht}. Using the large-$N$ results for the scaling dimension $\Delta_\psi$ and $\Delta_{\sigma_T}$, we can identify the bottom two kinks in Figure~\ref{fig:univ-bounds-sigmaT-zoomIn} as GNY models with $N=10, 20$. For lower values of $N$, it is plausible to conjecture that the other kinks in Figure~\ref{fig:univ-bounds-sigmaT-zoomIn} correspond to GNY models as well. Using this as a conjecture, Figure~\ref{fig:univ-bounds-sigmaT-zoomIn} then gives a non-perturbative estimate of $\Delta_\psi$ and $\Delta_{\sigma_T}$ in the GNY models at all $N$.

While it is satisfying to see that the GNY-model may saturate the universal bounds in the space of allowed scaling dimensions for this sector, we would want to determine the CFT data for operators in all the sectors of the low-lying spectrum. In order to learn about more of the spectrum we will however need to restrict the space of CFTs that we study. There are two possibilities in the present context: we can either assume gaps for the scaling dimension of operators in certain $O(N)$ representations and obtain tighter bounds, or, using the extremal functional method~\cite{Poland:2010wg,ElShowk:2012hu}, we can reconstruct the spectra of theories close to the kinks seen in Figure~\ref{fig:univ-bounds-sigmaT-zoomIn}. In the present work we will mostly focus on the former strategy, though we will mention some preliminary results from studying extremal functionals below.

As an example, we impose gaps in the $O(N)$ singlet parity-odd scalar sector of the theory. We will thus assume the existence of an operator in this sector with dimension $\Delta_\sigma$ and assume a gap above it until the operator $\sigma'$ with the dimension $\Delta_{\sigma'}$. We aim here to make a plot analogous to Figure~3 of Ref.~\cite{Iliesiu:2015qra}. There we did not impose $O(N)$ symmetry and used the gap to the second parity odd scalar (which corresponds to $\sigma_T$) as a proxy for $N$. For the each value of the gap, we observed a kink coinciding with the corresponding GNY values for $\Delta_\psi$ and $\Delta_\sigma$. In the present work, with the $O(N)$ symmetry imposed, we have the possibility to find a stable GNY kink for a range of gap assumptions. As long as the assumed gap on $\Delta_{\sigma'}$ is not larger than the correct value in the GNY model the kink can exist, and it should disappear only once we choose $\Delta_{\sigma'}$ too large.

Figure~\ref{fig:boundsWhenImposingGaps} explicitly shows the consequence of imposing a gap on $\Delta_{\sigma'}$: we plot the allowed region in the space of scaling dimensions $(\Delta_\psi$, $\Delta_\sigma)$ when imposing the gap $\Delta_{\sigma'} > 3$. This assumption certainly holds for large-$N$ GNY models, where, as shown in Table~\ref{tab:grossNeveuDimensions},  $\Delta_{\sigma'} = 3 + 64/(\pi^2 N) + \ldots$, but is not a priori justified for small values of $N$.  We take it as a working hypothesis, as it has an appealing interpretation that there is only one relevant scalar in this particular sector. 
By imposing the gap we carve out the allowed region below the free theory revealing new smoothed out ``kinks" on the boundary of the allowed region for the scaling dimension for each value of $N$. At small values of $N$, for each value, we observe the existence of two distinct kinks (besides that corresponding to the free theory, in the upper left corner).  We will first comment on the association of the top set of kinks with the GNY models, and then discuss the connection between the bottom set of kinks and the GNY$^*$ models.

For the set of kinks with higher values of $\Delta_\sigma$, the position of this feature once again sits near the value predicted by the large-$N$ expansion for the GNY theory: e.g. for $N=10$, indicated in Figure~\ref{fig:boundsWhenImposingGaps} through a black square. We conjecture that for lower values of $N$ the kinks can be used to read off the scaling dimensions $\Delta_\psi$ and $\Delta_\sigma$ for the GNY-model at strong coupling. This conjecture is further supported by results from the $\epsilon$-expansion (shown by the red shapes) whose estimates for the scaling dimensions $\Delta_\psi$ and $\Delta_\sigma$ are somewhat close to the top set of kinks even for small values of $N$.

\begin{table}[!t]
\centering
\begin{tabular}{c c c c c c c}
\hline\hline
 \hspace{0.6cm} $N$ \hspace{0.6cm} & \hspace{0.5cm} $2$ \hspace{0.5cm} & \hspace{0.5cm} $3$  \hspace{0.5cm} &   \hspace{0.5cm} $4$ \hspace{0.5cm}  & \hspace{0.5cm}8 \hspace{0.5cm} & \hspace{0.5cm}10 \hspace{0.5cm} &   \hspace{0.5cm} $20$ \hspace{0.5cm} \\
 \hline 
\hspace{0.6cm} $\Delta_\psi$ \hspace{0.6cm} & 1.067 & 1.054 & 1.042 & 1.021 & 1.017 & 1.008\\
\hspace{0.6cm} $\Delta_\sigma$ \hspace{0.6cm} & 0.660 & 0.724 & 0.772 & 0.871& 0.898& 0.944 \\
\hspace{0.6cm} $\Delta_{\sigma_T}$ \hspace{0.6cm} & 2.445 & 2.358 & 2.293& 2.153 & 2.121& 2.059\\
\hspace{0.6cm} $\Delta_\epsilon$ \hspace{0.6cm} &2.14  & 2.17 & 2.25 & 2.12 & 2.09 & 2.03\\
\hspace{0.6cm} $\Delta_{\sigma^\prime}$ \hspace{0.6cm} & 3.02 & 3.09 & 3.32& 3.52 & 3.59 & 3.39\\
\hspace{0.6cm} $\Delta_{\epsilon_T}$ \hspace{0.6cm} &  3.49 & 3.47 & 3.45& 3.29 & 3.24 & 3.14\\
\hline\hline
\end{tabular}\\\vspace{0.5cm}
 \caption{ 
Dimensions of some of the low-dimensional operators in GNY models obtained in this work. Dimensions of $\psi$, $\sigma$ and $\sigma_T$ can be obtained from Figures~\ref{fig:univ-bounds-sigmaT-zoomIn} and \ref{fig:boundsWhenImposingGaps} directly. Other dimensions were obtained from zeros of the extremal functionals and are thus given to less precision.}
\label{tab:grossNeveuDimensionsBootstrap}
\end{table}

\begin{table}[!t]
\centering
\begin{tabular}{c || c c c c}
\hline\hline
& \hspace{0.1cm} Conf.~boot.  \hspace{0.1cm} & \hspace{0.1cm} $\epsilon$-exp.  \hspace{0.1cm} &   \hspace{0.1cm} Func.~RG \hspace{0.1cm}  &   \hspace{0.1cm}Monte Carlo \hspace{0.1cm} \\
 \hline \hline
$N=1$  &  & & & \\
$\eta_{\psi}$ & 0.164\cite{Iliesiu:2015qra} & 0.162\cite{2017arXiv170308801M}, 0.176\cite{fei2016yukawa} & 0.180\cite{Heilmann:2014iga} & -  \\$\eta_{\sigma}$ & 0.164\cite{Iliesiu:2015qra} & 0.162\cite{2017arXiv170308801M}, 0.176\cite{fei2016yukawa}  & 0.180\cite{Heilmann:2014iga} &  - \\  
$1/\nu$& - & 1.419\cite{2017arXiv170308801M}, 1.412\cite{fei2016yukawa} & 1.408\cite{Heilmann:2014iga} & - \\\hline
$N=2$ & & & & \\
$\eta_{\psi}$ & 0.134 & 0.137\cite{2017arXiv170308801M} &0.112\cite{hofling2002phase} & - \\
$\eta_{\sigma}$ & 0.320 & 0.282\cite{2017arXiv170308801M}& 0.550\cite{hofling2002phase}& -\\
 $1/\nu$ & 0.86 &  1.493\cite{2017arXiv170308801M} & 1.614\cite{hofling2002phase} & -\\\hline
$N=4$ & & & & \\
$\eta_{\psi}$ & 0.084 & 0.102\cite{2017arXiv170308801M}, 0.096\cite{fei2016yukawa} &0.0645\cite{knorr2016ising} & - \\
$\eta_{\sigma}$ & 0.544 & 0.463\cite{2017arXiv170308801M}, 0.506\cite{fei2016yukawa}& 0.550\cite{knorr2016ising}& 0.45(3)\cite{li2015fermion} \\
$1/\nu$ & 0.76 & 1.166\cite{2017arXiv170308801M}, 0.852\cite{fei2016yukawa} & 1.075(4)\cite{knorr2016ising} & 1.30(5)\cite{li2015fermion}  \\
\hline
$N=8$  & & & &  \\
$\eta_{\psi}$ & 0.044 & 0.074\cite{2017arXiv170308801M}, 0.082\cite{Gracey:2016mio}, 0.042\cite{fei2016yukawa} & 0.0276\cite{knorr2016ising}  & 0.38(1)\cite{chandrasekharan2013quantum} \\
$\eta_{\sigma}$ & 0.742 & 0.672\cite{2017arXiv170308801M}, 0.745\cite{Gracey:2016mio}, 0.74\cite{fei2016yukawa} & 0.7765\cite{knorr2016ising} & 0.62(1)\cite{chandrasekharan2013quantum} \\
$1/\nu$ & 0.88 & 1.048\cite{2017arXiv170308801M},  0.931\cite{Gracey:2016mio}, 0.948\cite{fei2016yukawa} &0.994(2)\cite{knorr2016ising} & 1.20(1)\cite{chandrasekharan2013quantum}  \\\hline
\hline
\end{tabular}\\\vspace{0.5cm}
 \caption{ Anomalous dimensions, $\eta_\sigma = 2\left({\Delta_\sigma} - \frac{1}2\right)$ and $\eta_\psi = 2({\Delta_\psi - 1})$, and  correlation length exponent, $1/\nu = 3-\Delta_\epsilon$, for $N=1,\, 2,\, 4,\, 8$. Conformal bootstrap results for $N=1$ are taken from our previous work in \cite{Iliesiu:2015qra} by intersecting the curve of kink found by imposing a gap on the dimension $\Delta_{\sigma'}$ with the SUSY line $\Delta_{\psi} = \Delta_\sigma +1/2$. The bootstrap results for $1/\nu$ are obtained from estimating  $\Delta_\epsilon$ using the extremal functional method. The results shown from \cite{2017arXiv170308801M} for the anomalous dimensions and for the correlation length exponent  are the Pad\'e$_{[2, 1]}$ approximations obtained from the $\epsilon$-expansion at three loops (see Eqs.~(11)--(13) and Table II in \cite{2017arXiv170308801M}). The results shown from \cite{fei2016yukawa} for the same exponents are the two-sided Pad\'e$_{[4, 2]}$,  Pad\'e$_{[4, 2]}$ and, respectively, the Pad\'e$_{[1, 5]}$ obtained from the two-loop $2+\epsilon$ and $4-\epsilon$ expansion (see Table 1 in \cite{fei2016yukawa}). The results from \cite{Gracey:2016mio} come from the $\epsilon$-expansion at four loops with no Pad\'e resummation performed (see Table 2 in \cite{Gracey:2016mio}).}
\label{tab:exp-relev}
\end{table}
Interestingly, the $\epsilon$-expansion also provides a possible explanation for the bottom set of kinks in Figure~\ref{fig:boundsWhenImposingGaps}: they may be alternative fixed-points, dubbed GNY$^*$ models. Specifically, the values for the scaling dimensions $\Delta_\psi$ and $\Delta_\sigma$ estimated by the  $\epsilon$-expansion expansion, shown by the green shapes, are close to the lower set of kinks.\footnote{The three-loop $\epsilon$-expansion Pad\'e estimates for GNY and GNY$^*$ were kindly provided by Grigory Tarnopolsky in private  discussions, using the methods presented in \cite{fei2016yukawa} and \cite{2017arXiv170308801M}.}
While this is encouraging,  our assumption that $\Delta_{\sigma'} > 3$ is disfavored by the $\epsilon$-expansion results for the GNY$^*$ theories which suggest, for instance for $N=4$, the estimate $\Delta_{\sigma'}\approx 1.9$.\footnote{However, preliminary investigations of extremal functionals suggest the lower set of kinks may persist down to $\Delta_{\sigma'}\approx 2.5$.} It will be important to perform a bootstrap study better tailored to the GNY$^*$ models in the future.

Thus, by imposing a minimal set of gaps, we find kinks corresponding to the GNY models that give information about some operator dimensions in those theories. However, we can potentially study all sectors of the GNY model using the extremal functional method \cite{Poland:2010wg,ElShowk:2012hu,El-Showk:2014dwa,Simmons-Duffin:2016wlq}. To be effective, the extremal functional method requires a point on the boundary of the allowed region of CFT data that is close to the theory of interest. (For example, because the bounds on the Ising model are so strong, the extremal functional method gives good estimates for several operators in that theory \cite{El-Showk:2014dwa,Simmons-Duffin:2016wlq}.) We have performed a preliminary study of the extremal functionals in the case of GNY models. The resulting spectra are somewhat noisy, but they allow an estimate of low dimension scalar operators, which we have included in Table~\ref{tab:grossNeveuDimensionsBootstrap}. 

In Table~\ref{tab:exp-relev}, we give a comparison of critical exponents obtained from our bootstrap study as well as results from the $\epsilon$-expansion, the functional RG method, and Monte Carlo simulations. (The relation between the anomalous dimensions $\eta_\sigma$ and $\eta_\psi$, the correlation length exponent $\nu^{-1}$, and the scaling dimensions computed in this paper are given by
\be
\Delta_{\sigma}= 1-\frac{\epsilon}2 + 
\frac{\eta_\sigma}2\,, \text{\hspace{1cm}}\Delta_{\psi} = \frac{3}2 - \frac{\epsilon}2 + \frac{\eta_\psi}2\,, \text{\hspace{1cm}}\Delta_\epsilon = 4 - \epsilon - \frac{1}\nu \,,
\ee
where one should set $\epsilon = 1$ in order to extrapolate to three dimensions.)   We note that our results for the anomalous dimensions are closest to some of the $\epsilon$-expansion Pad\'e approximations performed in \cite{fei2016yukawa}.  However, both our bootstrap results and the perturbative results differ significantly from the Monte Carlo simulations.

\section{Universal Bounds on Scaling Dimensions }
\label{sec:generic-bounds}
In this section we present additional general bounds on low-lying scalars in the $\psi_i\times\psi_j$ OPE. We consider scalars of different parities in various representations of $O(N)$. We explore the bounds at somewhat higher values of the fermion dimension $\Delta_\psi$ than in Section~\ref{sec:GNY_spectrum}. The results exhibit some unusual jumps reminiscent of the features previously seen in the fermion bootstrap of~\cite{Iliesiu:2015qra} where no global symmetries were imposed.

\subsection{The Lowest Dimension Parity-Odd Scalar Singlet}
\label{sec:bound-parity-odd-scalar-singlet}

 \begin{figure}[t!]
    \centering
    Upper bounds on scaling dimension of lowest-lying parity-odd singlet, $\Delta_{\sigma}$
    
\includegraphics[width=0.85\textwidth]{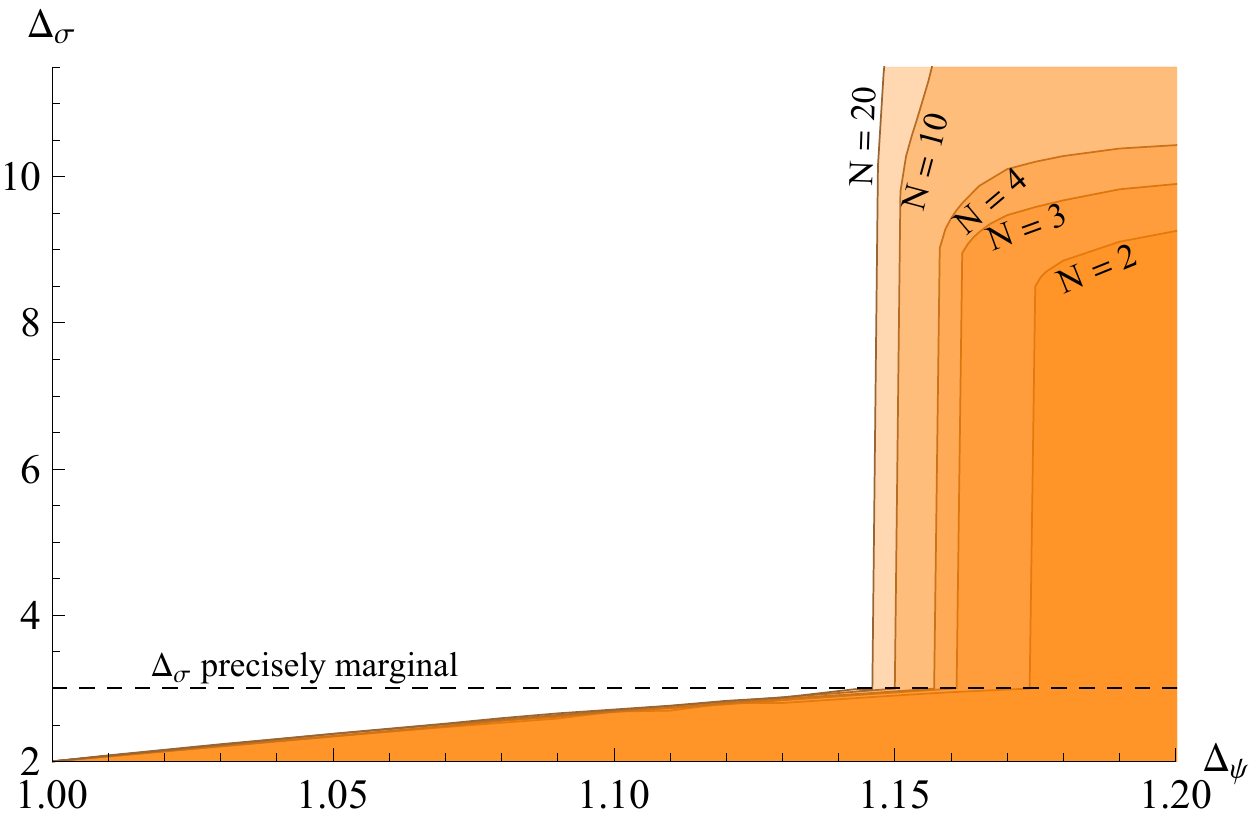}
    \caption{\label{fig:univ-bounds-sigma}Upper bounds on the scaling dimension of the lowest-lying parity-odd $O(N)$ singlet, $\Delta_{\sigma}$, for a unitary CFT containing fermions with dimension $\Delta_\psi$. Above, we focus on the cases $N=2,\, 3,\,4, \, 10$ and $20$.  As $\Delta_\psi \to 1$, the bound approaches the free theory value of $\Delta_{\sigma} = 2$. Reminiscent of the jump noticed in our previous work \cite{Iliesiu:2015qra} when bounding the scaling dimension of the lowest-lying parity-odd operator in a theory with one fermion, we observe a different jump for each value of $N$ that we study. Once again, all jumps occur once the bound intersects the horizontal line on which $\sigma$ is precisely marginal, $\Delta_{\sigma} = 3$. }
  \end{figure}
  
In Figure~\ref{fig:univ-bounds-sigma}, we plot universal upper bounds on the scaling dimension of the lowest-lying parity-odd scalar singlet $\Delta_{\sigma}$ as a function of $\Delta_{\psi}$, for any 3D parity-invariant CFT with a global $O(N)$ symmetry. The bound starts at the point $(\Delta_\psi, \Delta_{\sigma})=(1,2) $, corresponding to the free theory with $N$ fermions. For each value of $N$, the bound then increases monotonically up to the point where it intersects the horizontal line $\Delta_{\sigma} = 3$ (corresponding to a marginal operator). At these intersection points, a sharp vertical discontinuity occurs, after which the bounds plateau at much higher values of $\Delta_{\sigma}$. For instance, at $N=2$ we find a plateau around $\Delta_{\sigma}\approx 9.5$, while for $N=10$ we find $\Delta_\sigma\approx 16$. The intersection points occur at lower values of $\Delta_{\psi}$ for larger $N$, while the jump in $\Delta_{\sigma}$ increases with $N$. 

From these bounds, we can at least determine the values of $\Delta_{\psi} $ for which the CFTs must have a relevant parity-odd singlet in the $\psi_i \times \psi_j$ OPE\@. For instance, for $N=2$ we conclude that 3D parity invariant CFTs with an $O(2)$ global symmetry that have $\Delta_{\psi} < 1.175$ must also have at least one relevant parity-odd singlet scalar. Conversely, all such theories that have no relevant parity-odd singlet scalar must have $\Delta_{\psi} > 1.175$.

The jumps shown in Figure~\ref{fig:univ-bounds-sigma} are reminiscent of features previously encountered in studies of fermions without a continuous global symmetry~\cite{Iliesiu:2015qra}. In Figure~1 from \cite{Iliesiu:2015qra}, we observed that when  $\Delta_\psi \approx 1.27$, the bound for the dimension of the lowest-lying parity-odd scalar jumps from $\Delta_{\sigma} \approx 3$ to $\Delta_{\sigma} \approx 7.7$. When bounding dimensions in the parity-even sector for such theories, one finds a kink at the same value of $\Delta_\psi$ with the scaling of the lowest-lying parity-even scalar $\Delta_\epsilon \approx 5.1$. As noted in \cite{Iliesiu:2015qra}, the observed jump is similar to ones corresponding to the 3D Ising model, when studying scalar dimension bounds using mixed-correlators \cite{Kos:2014bka}.  By analogy with the existence of the 3D Ising model at the location of the jumps, we therefore conjectured that there exists a ``dead-end'' parity-invariant CFT without any relevant scalar operators, with $\Delta_{\psi} \approx 1.27$ which would have very large anomalous dimensions in both the parity-even and odd sectors, $\Delta_{\epsilon} \approx 5.1$ and $3 < \Delta_\sigma < 7.7$. It is therefore tempting to extend the line of reasoning from \cite{Kos:2014bka} and conjecture the existence of a family of ``dead-end'' theories with an $O(N)$ global symmetry. For each value of $N$, the dimension of the scalar singlet in the parity-odd scalar in such a theory would satisfy $3 < \Delta_{\sigma} < \Delta_{\sigma}^{\text{max}}(N)$ where  $\Delta_{\sigma}^{\text{max}}(N)$ is the upper bound of each jump which is increasing monotonically with $N$. In the future it will be interesting, for instance, to study the extremal spectra of these jumps and try to find further evidence for the existence of new $O(N)$-symmetric ``dead-end" CFTs.

\subsection{The Lowest Dimension Parity-Even Scalar Singlet}
\label{sec:bound-parity-even-scalar-singlet}

   \begin{figure}[t!]
    \centering
    Upper bounds on scaling dimension of lowest-lying parity-even singlet, $\Delta_{\epsilon}$
    
\includegraphics[width=0.85\textwidth]{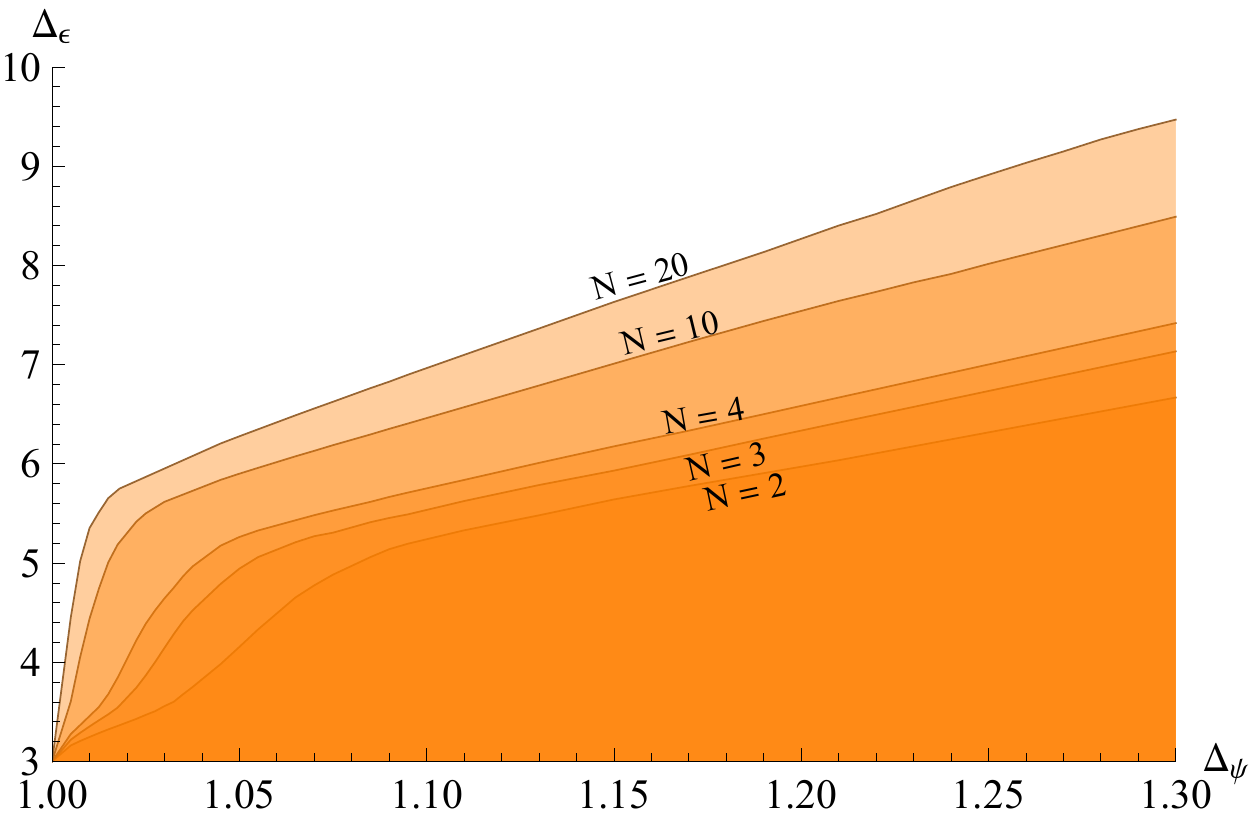}
    \caption{\label{fig:univ-bounds-epsilon}Upper bounds on the scaling dimension of the lowest-lying parity-even $O(N)$ singlet, $\Delta_{\epsilon}$, for a unitary CFT containing fermions with scaling dimension $\Delta_\psi$. Once again, we focus on the cases $N=2,\, 3,\,4, \, 10$ and $20$.  As $\Delta_\psi \to 1$, the bound approaches the value of $\Delta_{\epsilon} = 3$. }
  \end{figure}

We now focus on the parity-even singlet sector.  Figure~\ref{fig:univ-bounds-epsilon} shows an upper bound on the scaling dimension of the lowest-lying operator in this sector as a function of $\Delta_{\psi}$.  The bound starts at $(\Delta_{\psi}, \Delta_\epsilon) = (1, 3)$,\footnote{This places the free theory, for which the lowest-lying parity-even singlet scalar has $\Delta_\epsilon = 2$, in the allowed region.} and increases monotonically as $\Delta_{\psi}$ is increased, with an inflection point that gets closer to $\Delta_{\psi}\rightarrow 1$ as $N$ is increased. In~\cite{Iliesiu:2015qra}, the jump in the parity odd sector coincided with a kink at the same value of $\Delta_\psi$ in the parity even sector. However, this is not the case for the bounds presented in Figure~\ref{fig:univ-bounds-epsilon}.  Thus, if the conjectured family of ``dead-end'' CFTs from Section~\ref{sec:bound-parity-odd-scalar-singlet} were to exist, they should lie within the allowed region, for instance implying that the theories with $O(2)$ symmetry should have $\Delta_{\epsilon} < 5.7$.

Interestingly, the values of $\Delta_{\psi}$ at which we find the inflection point for the boundary curve from Figure~\ref{fig:univ-bounds-epsilon} coincidentally seems to match the value predicted by the large-$N$ expansion for the GNY-model. However, as can be seen in top of Table~\ref{tab:grossNeveuDimensions}, the value predicted for the dimension of the parity-even scalar $\Delta_\epsilon$, which is identified with the scaling dimension of $\phi^2$, is well within the bootstrap-determined universal bounds. 

Thus, the connection between these features and the GNY models is uncertain. However, as discussed in the next section, when placing universal bounds on the scaling dimension of the parity-odd symmetric traceless scalar, the presence of the GNY model on the boundary becomes apparent.

\subsection{The Lowest Dimension Parity-Odd Symmetric Traceless Scalar}

  \begin{figure}[t!]
    \centering
    Upper bounds on scaling dimension of lowest-lying parity-odd traceless symmetric operator, $\Delta_{\sigma_T}$
    
\includegraphics[width=0.85\textwidth]{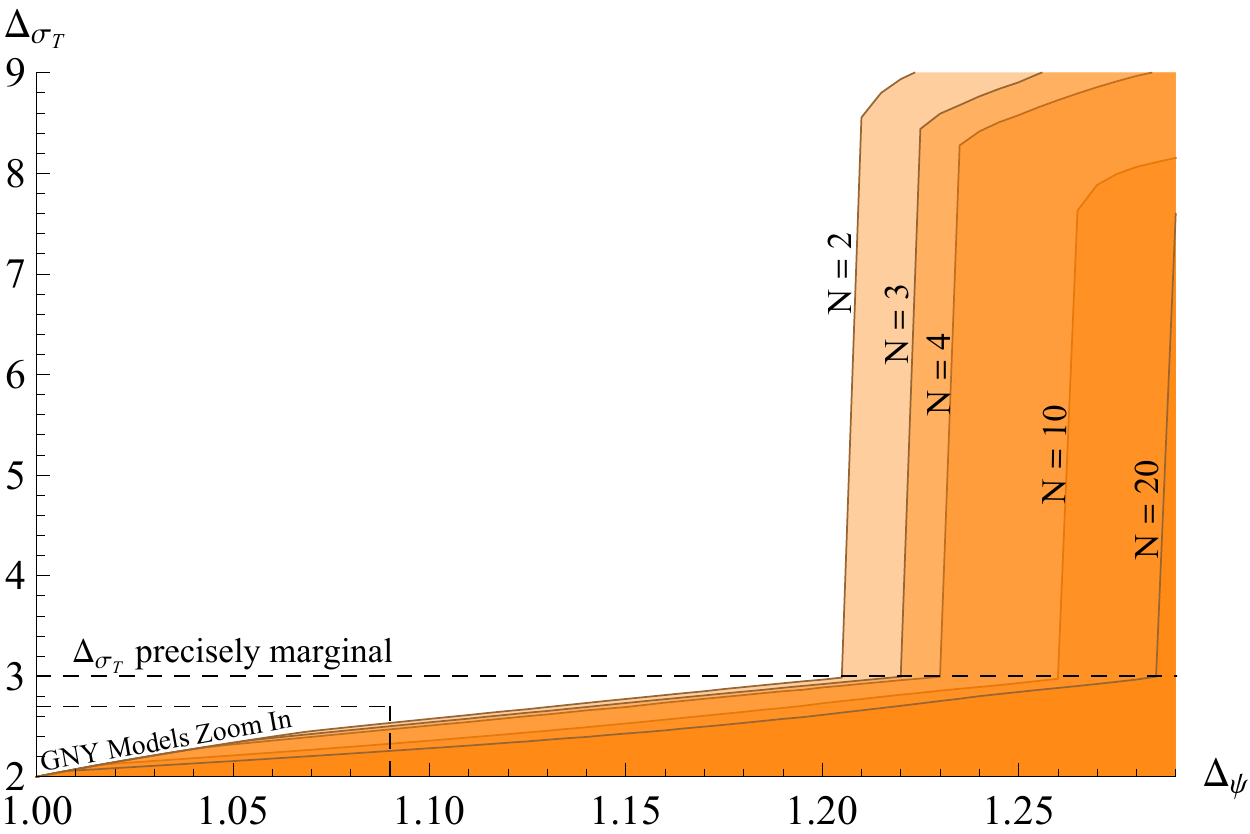}
        \caption{\label{fig:univ-bounds-sigmaT}Upper bounds on the scaling dimension of the lowest-lying parity-odd $O(N)$ symmetric-traceless tensor, $\Delta_{\sigma_T}$, for a unitary CFT containing fermions with scaling dimension $\Delta_\psi$. Once again, we focus on the cases $N=2,\, 3,\,4, \, 10$ and $20$. As $\Delta_\psi \to 1$, the bound approaches the free theory value of $\Delta_{\sigma_T} = 3$.  There are two sets of kinks. The first set of kinks correspond to the GNY-models and are located in the square on the lower left. We zoom onto this lower left square in Figure~\ref{fig:univ-bounds-sigmaT-zoomIn} and discuss the meaning of the kinks in more detail in Section~\ref{sec:GNY_spectrum}. }
  \end{figure}

In Figure~\ref{fig:univ-bounds-sigmaT} we show an upper bound for the scaling dimension of the lowest-lying parity-odd $O(N)$ symmetric traceless scalar, $\Delta_{\sigma_T}$, as a function of $\Delta_\psi$. Once again, the bound starts at the point corresponding to the free theory with  $(\Delta_\psi, \Delta_{\sigma_T}) = (1,2)$. For all values of $N$, each bound increases monotonically with $\Delta_\psi$, encountering a first kink at points with values of $\Delta_{\psi}$ and $\Delta_{\sigma_T}$ ranging from $(\Delta_{\psi}, \Delta_{\sigma_T}) \approx (1.069, \,2.45)$ for $N=2$, to $(\Delta_{\psi}, \Delta_{\sigma_T}) \approx (1.008,\, 2.06)$ for $N=20$. The location of these kinks can be seen more clearly in Figure~\ref{fig:univ-bounds-sigmaT-zoomIn}, where we zoom in on the region of interest from Figure~\ref{fig:univ-bounds-sigmaT}.  For large values of $N$ we find a strong agreement between the position of the kinks and large-$N$ expansion estimates for the values of the scaling dimensions, $\Delta_{\psi}$ and $\Delta_{\sigma_T}$, in GNY-theories. 

As we move towards larger values of $\Delta_\psi$ we find yet another discontinuity: for all values of $N$ we once again find a jump in the bound on $\Delta_{\sigma_T}$ that occurs precisely where the bound intersects the horizontal line where $\sigma_T$ is precisely marginal. The value of $\Delta_{\psi}$ at which the jump occurs increases monotonically with $N$, ranging from $\Delta_{\psi} \approx 1.21$ for $N=2$ to $\Delta_{\psi}\approx 1.28 $ for $N =20$.  The range of the jump in the value of $\Delta_{\sigma_T}$ decreases with $N$, as for $N=2$ one finds $\Delta_{\sigma_T} \leq 8.6$, while for the highest value of $N$ ($N=20$) that we have used numerics for we find  $\Delta_{\sigma_T} \leq 7.2$.

Once again following the reasoning presented in Section~\ref{sec:bound-parity-odd-scalar-singlet} as well as in our previous work \cite{Iliesiu:2015qra}, we are led to consider the possibility that the set of jumps follows a second  family of theories, namely a second extension of the ``dead-end'' CFT seen when placing universal bounds on fermionic theories with no $O(N)$ global symmetry.

While we cannot yet check the existence of these conjectured CFTs, Figure~\ref{fig:univ-bounds-sigmaT} allows us to conservatively claim for what values of $\Delta_{\psi}$ there must be at least one relevant operator in this sector: for instance, for $N=2$ we find that if $\Delta_{\psi} < 1.21$ then the theory needs at least one relevant operator in the symmetric traceless representation, and conversely, if we want to study CFTs with no relevant operators in this sector, we need $\Delta_{\psi} > 1.21$.

\section{Universal Bounds on Central Charges}
\label{sec:ccbounds}

In this section we place lower bounds on the central charge of the $O(N)$ conserved current $C_J$, and on the stress-energy tensor central charge $C_T$. These quantities are defined as coefficients in the the two-point function of the corresponding currents,
\begin{align}
  \langle J^\mu_{\text{can}, ij}(x_1) J^\nu_{\text{can}, kl}(x_2) \rangle &= (\delta_{ik}\delta_{jl} - \delta_{il}\delta_{jk}) \frac{C_J}{(4\pi)^2}
   \frac{ I^{\mu\nu}(x_{12})}{x_{12}^{4}} \,,\label{eq:norm-CC-cons-current} \\
      \langle T_\text{can}^{\mu \nu}(x_1) T_\text{can}^{\rho \sigma} (x_2)\rangle &= \dfrac{C_T}{(4\pi)^2} \dfrac{1}{x_{12}^6} \left[ \frac 12 \left( I^{\mu\rho}(x_{12}) I^{\nu\sigma}(x_{12}) + I^{\mu\sigma}(x_{12}) I^{\nu\rho}(x_{12})\right) - \frac 13 \eta^{\mu\nu} \eta^{\rho\sigma}   \right]\,,\label{eq:norm-CC-stress-tensor}
\end{align}
with $I^{\mu \nu} (x) \equiv \eta^{\mu\nu} - 2 x^\mu x^\nu / x^2$. Here $J^\mu_{\text{can}}$ and $T^{\mu \nu}_{\text{can}}$ denote the canonically normalized $O(N)$ conserved current and stress-energy tensor, defined such that a theory of free $N=1$ Majorana fermions has $C_J^\text{free} = 2$ and $C_T^\text{free} = 1$. The conserved current $J^\mu_{\text{can}}$ is the spin-1, parity-even, antisymmetric operator with dimension $\Delta_J = 2$, while the operator $T^{\mu \nu}_{\text{can}}$ is the spin-2, parity-even, $O(N)$ singlet operator with scaling dimension $\Delta_T = 3$.

\subsection{Bounds on the $O(N)$ Current Central Charge}

Let us first determine how $C_J$ appears in the conformal block decomposition.
We start by using the Ward identity for $O(N)$-transformations,
\begin{align}
\frac{\partial}{\partial x^\mu}  \langle J_{\text{can}, ij}^{\mu}(x) \psi_k(x_1)& \psi_l(x_2) \rangle  + \delta(x - x_1)( -\delta_{ik} \langle \psi_j(x_1) \psi_l(x_2) \rangle +\delta_{jk} \langle \psi_i(x_1) \psi_l(x_2) \rangle ) \notag\\
&+\delta(x - x_2)( -\delta_{il} \langle \psi_k(x_1) \psi_j(x_2) \rangle +\delta_{jl} \langle \psi_k(x_1) \psi_i(x_2) \rangle ) = 0 \,, 
\end{align}
where  $J^{\mu}_{\text{can}, ij}$ denotes the canonically normalized conserved current (\ref{eq:norm-CC-cons-current}), antisymmetric in the indices $i$ and $j$. From this Ward identity, we find a condition on the OPE coefficients $\lambda^a_{J, \text{can}}$ of the current $J^{\mu \nu}_{\text{can}, ij}$ appearing in the $\psi_i \times \psi_j$ OPE:
 \es{GotlambdaJMainText}{
   \mu_{ J,  \text{can}}^1 \equiv 2 \lambda_{ J,  \text{can}}^1 - \lambda_{J, \text{can}}^2 = - \frac{i}{2 \pi} \,.
 }
 Here we introduced a new basis for 3-point functions and corresponding OPE coefficients $\mu^a$:
 \es{eq:mudefinition}{
 \mu^1 \equiv 2 \lambda^1 - \lambda^2\,, \qquad  \mu^2 \equiv \lambda^1 + 2 \lambda^2\,.
 }
 
In our bootstrap setup, the normalization of operators appearing in the $\psi_i \times \psi_j$ OPE depends only on $\Delta$ and $\ell$ and is otherwise independent of the details of the CFT we study. Thus, we need to relate the canonically normalized current to the operators we use in our setup, which have normalization fixed by
\be
\label{eq:stand-norm}
  \langle \cO_\ell(x_1, s_1) \cO_\ell(x_2, s_2) \rangle = i^{2\ell}  \frac{ (s_1 (x_1-x_2) s_2 )^{2\ell}}{x_{12}^{2\Delta_{\cO} + 2\ell}}  \,.
\ee
 The comparison with (\ref{eq:norm-CC-cons-current}) gives:
\es{JRelation}{
  J^{\mu} = \frac{4 \pi}{\sqrt{2C_J}} J^{\mu}_\text{can}, \quad \text{ and consequently,}\quad
   \lambda_{J}^{1,2} =\lambda_{J, \text{can}}^{1,2} \frac{4\pi}{\sqrt{2 C_J}}\,, \quad  \mu_{J}^{1,2} =\mu_{J, \text{can}}^{1,2} \frac{4\pi}{\sqrt{2C_J}}\,.
}
We can now rearrange the terms in the crossing equation as follows:
\begin{align}
\left( \mu^1_{J} \right)^2 &\hat {\vec V}^{I, A}_{11,2, 1} + \mu^1_{J} \mu^2_{J} \left[ \hat {\vec V}^{I, A}_{12,2, 1} + \hat {\vec V}^{I, A}_{21,2, 1} \right]= -\sum_{a,b=1,2}\l_{\mathbb{1}}^a\l_{\mathbb{1}}^b \vec V_{ab,0,0}^{I, S} -  \left( \mu^2_{J} \right)^2 \hat {\vec V}^{I, A}_{22,2, 1}  \notag\\
&- \sum_{\substack{\cO^+_R, \,\ell \in \ell^R\,\textrm{even}\\a,b=1,2}} \lambda^a_{\cO^+} \lambda^b_{\cO^+} \vec V^{I, R}_{ab,\De,\ell} - \sum_{\cO^-_R, \,\ell\in \ell^R} (\lambda^3_{\cO^-_R})^2  \vec V^{I, R}_{33,\De,\ell} - \sum_{\cO^-_R, \,\ell\in \bar\ell^R} (\lambda^4_{\cO^-_R})^2 \vec V^{I, R}_{44,\De,\ell}\,, \label{eq:crossingcj}
\end{align}
 where we suppressed the $u,v$-dependence and used the hat on $\vec V$ to indicate that we switched to a 3-point function basis corresponding to $\mu^a$'s. We now search for functionals $\vec \alpha_i$  satisfying conditions (\ref{eq:constraintsonfunctionalforcentralcharge}). Notice, however, that only $\mu^1_{J,\text{can}}$ is fixed by Ward identity, while $\mu^2_{J,\text{can}}$ is at this stage arbitrary. To place the bound on $\mu_J^1$ (equivalently, $C_J$) irrespectively of the value of $\mu_J^2$, and to avoid the costly scanning over all possible values of $\mu_J^2$, we look for the functionals which satisfy the additional condition:
\be
\vec \alpha_I \cdot \left[ \hat {\vec V}^{I, A}_{12,2, 1} + \hat {\vec V}^{I, A}_{21,2, 1} \right] = 0\,.
\ee
We can now proceed in exactly the same way as described in Section~\ref{sec:crossing} to put a lower bound on $C_J$.

   \begin{figure}[t!]
    \centering
    Lower bounds on the $O(N)$ current central charge $C_J$
    
\includegraphics[width=0.85\textwidth]{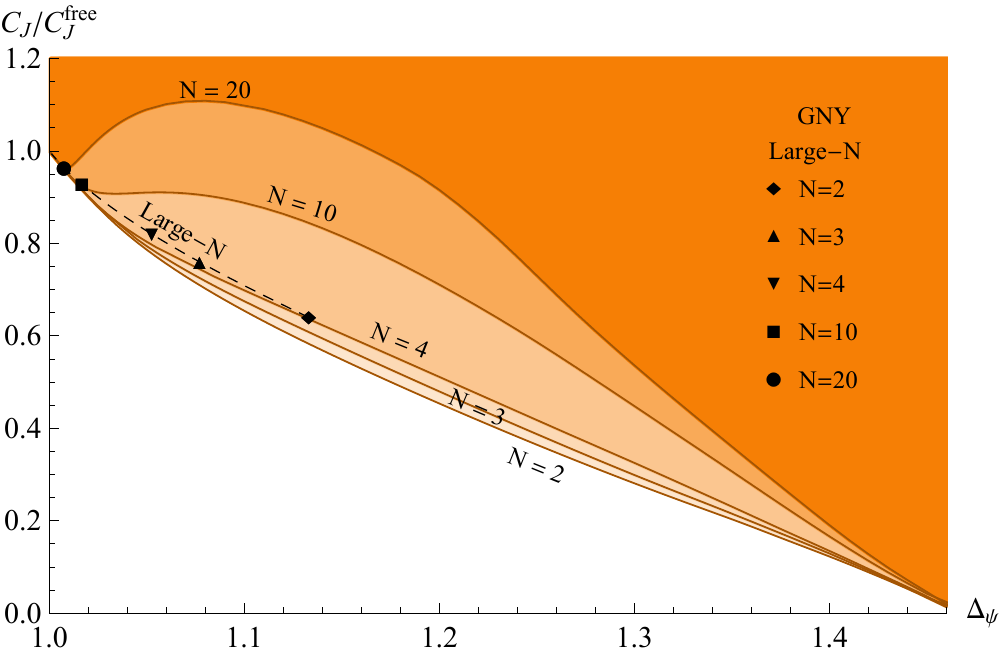}
          \caption{\label{fig:centralCharge-Cons-Curr} Lower bounds on the $O(N)$ conserved current central charges $C_J$ as a function of the scaling dimension $\Delta_\psi$. We normalize the central charges $C_J$ such that as $\Delta_{\psi} \rightarrow 1$, we set the value of $C_J=1$. For $N=10, 20$ the bound has a local minimum that lies close to the large-$N$ values of $(\Delta_\psi, C_J)$ in GNY models obtained up to  $O(1/N^2)$ corrections (black markers) \cite{diab2016c_j}. This is reminiscent of the central charge bound in the scalar bootstrap which had a local minimum corresponding to 3D Ising model. The bounds for smaller values of $N$ have no similar features.  }
  \end{figure}

Figure~\ref{fig:centralCharge-Cons-Curr} shows universal lower bounds for the current central charge $C_J$ in the units of free field value $C_J^\text{free}=2$, as a function of the fermion scaling dimension $\Delta_\psi$. At the unitarity bound $\Delta_\psi =1$ the bound goes to $C_J=C_J^\text{free}$. In other words, the bound is saturated by the theory of $N$ free Majorana fermions for all values of $N$. For large values of $N$ the bound has a sharp local minimum at scaling dimensions $\Delta_\psi$ corresponding to GNY models. The values of $C_J$ at the minimum agree with the large-$N$ estimates of $C_J$ for $N=10, 20$. This is analogous to the phenomena observed in \cite{ElShowk:2012ht, El-Showk:2014dwa} where the Ising model was found lying in the local minimum of the $C_T$ bound, and similarly in \cite{Kos:2013tga} where the same was found for $O(N)$ vector models under certain assumptions.

For smaller values of $N$ we do not observe any local minima or even sharp changes in the slope of the bound. It is possible that increasing the derivative cutoff $\Lambda$ in our computations (see Appendix \ref{app:sdpb}) would improve the bounds significantly and result in local minima even for small values of $N$. For large $N$ at least, we can see that the GNY models saturate the bootstrap bound and furthermore lie at a special point (minimum) of the bound, similarly to the bounds on the scaling dimension $\Delta_{\sigma_T}$ in Section~\ref{sec:GNY_spectrum}. Note that we could try to use the functional obtained in minimization of $C_J$ to extract the spectrum of the GNY model in question. In our preliminary investigations we have found that functionals obtained from scaling dimension bounds seem to give more precise results.

\subsection{Bounds on the Stress-Energy Tensor Central Charge}

Using the Ward identity for translations,
 \es{StressWard}{
  \frac{\partial}{\partial x^\mu}  \langle T_\text{can}^{\mu\nu}(x) \cO_1(x_1) \ldots \cO_n(x_n) \rangle 
    + \sum_{i=1}^n \delta(x - x_i) \frac{\partial}{\partial x_i^\nu}
   \langle \cO_1(x_1) \ldots \cO_n(x_n) \rangle = 0 \,,
 }
one can determine the OPE coefficients $\lambda^a_{T, \text{can}}$ of the canonically normalized stress-energy tensor $T^{\mu \nu}_{\text{can}}$, appearing in the $\psi_i \times \psi_j$ OPE. Specifically, one finds \cite{Iliesiu:2015qra}
 \es{GotlambdaMainText}{
   \lambda_{ T, \text{can}}^1 = \frac{3 i (\Delta_\psi - 1)}{8 \pi} \,, \qquad
    \lambda_{T, \text{can}}^2 = -\frac{3 i}{4 \pi}  \,.
 }

Once again we must relate the canonically normalized stress-energy tensor to the one we use in our bootstrap program, written in Eq. (\ref{eq:stand-norm}), which is equivalent to imposing that the 2-point function of the stress-energy tensor is normalized as
\be
 \langle T^{\mu \nu}(x_1) T^{\rho \sigma} (x_2)\rangle = \dfrac{1}{4 x_{12}^6} \left[ \frac 12 \left( I^{\mu\rho}(x_{12}) I^{\nu\sigma}(x_{12}) + I^{\mu\sigma}(x_{12}) I^{\nu\rho}(x_{12})\right) - \frac 13 \eta^{\mu\nu} \eta^{\rho\sigma}   \right].
\ee
  From (\ref{eq:norm-CC-stress-tensor}) and the equation~above, one finds 
 \es{TRelation}{
  T^{\mu\nu} = \frac{2 \pi}{\sqrt{C_T}} T^{\mu\nu}_\text{can}, \qquad \text{ and consequently,}\qquad
   \lambda_{T}^{1,2} =\lambda_{T, \text{can}}^{1,2} \frac{2\pi}{\sqrt{C_T}}\,.
 }
 
We can now put a lower bound on $C_T$ by bounding the OPE operators $\lambda^{1,2}_T$. We do this by isolating the contribution of the parity-even spin-2 operator with $\Delta =3$ from the singlet sector in the crossing Eqs.~\eqref{eq:first-crossing-eq}--\eqref{eq:third-crossing-eq}, 
\begin{align}
\lambda^a_{T} \lambda^b_{T} \vec V^{I_{\pm}, S}_{ab,3, 2}(u,v) = &-\sum_{a,b=1,2}\l_{\mathbb{1}}^a\l_{\mathbb{1}}^b \vec V_{ab,0,0}^{I_\pm, S}(u,v)   - \sum_{\substack{\cO^+_R, \,\ell \in \ell^R\,\textrm{even}\\a,b=1,2}} \lambda^a_{\cO^+} \lambda^b_{\cO^+} \vec V^{I_{\pm}}_{ab,\De,\ell}(u,v) \notag\\
&- \sum_{\cO^-_R, \,\ell\in \ell^R} (\lambda^3_{\cO^-_R})^2  \vec V^{I_{\pm}, R}_{33,\De,\ell}(u,v) - \sum_{\cO^-_R, \,\ell\in \bar\ell^R} (\lambda^4_{\cO^-_R})^2 \vec V^{I_{\pm}, R}_{44,\De,\ell}(u,v)\,, \label{eq:crossingct}
\end{align}
where the summation in the second term on the right-hand side now excludes the stress energy tensor and the identity operator, whose contributions we wrote separately. We now search for the function $\vec \alpha_i$ such that the conditions (\ref{eq:constraintsonfunctionalforcentralcharge}) are satisfied and, at the same time, minimize $-\vec \alpha_i [\vec V_{11, 0, 0}^{I_\pm}(u, v)]$. 

    \begin{figure}[t!]
    \centering
    Lower bounds on the stress-energy tensor central charge $C_T$
    
\includegraphics[width=0.85\textwidth]{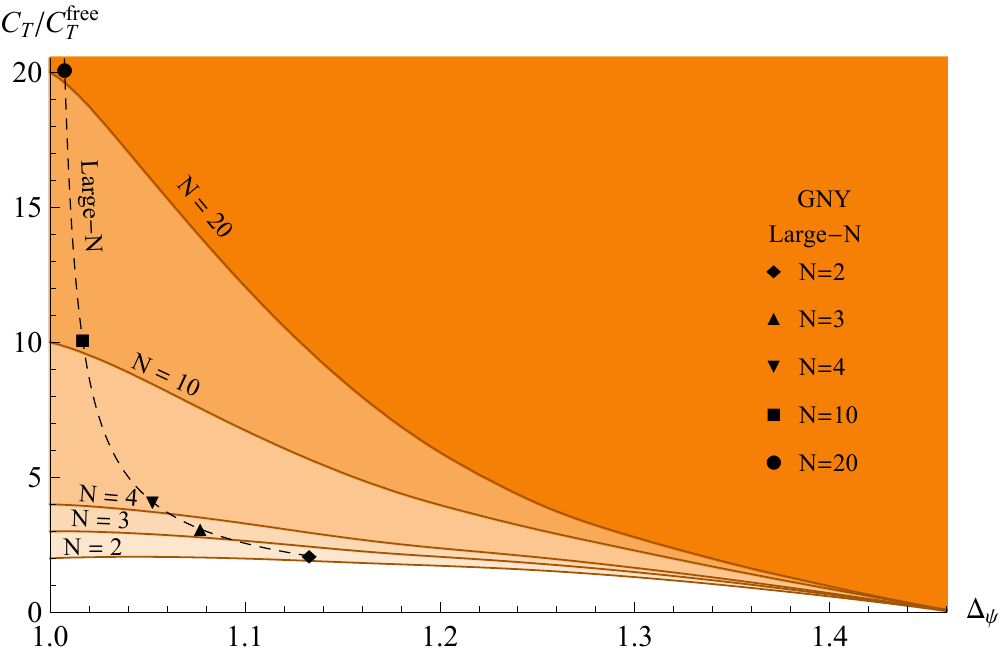}
        \caption{\label{fig:centralChargeStressTensor} Lower bounds on the stress energy tensor central charge $C_T$ as a function of the scaling dimension $\Delta_\psi$ of the fermion transforming in the $O(N)$  fundamental representation. We normalize the central charges $C_T$ such that as $\Delta_{\psi} \rightarrow 1$, we set the value of $C_T$ to be equal to the number of fermions in the theory. The black markers are large-$N$ estimates for GNY models obtained up to $O(1/N^2)$ corrections \cite{diab2016c_j}.  }
  \end{figure}

Figure~\ref{fig:centralChargeStressTensor} shows universal lower bounds on $C_T$ as a function of $\Delta_\psi$. For all values of $N$, the bound starts at $(\Delta_{\psi}, C_T) = (1, N)$, which is saturated by the free theory with $N$ Majorana fermions. The bound then decreases as a function of $\Delta_\psi$ until $\Delta_\psi \approx 1.465$, at which point all values of $C_T$ consistent with unitarity become allowed. While the GNY-model does not saturate the bounds in Figure~\ref{fig:centralChargeStressTensor}, it is important to check that the recently obtained large-$N$ estimate for the central charge $C_T$ is in the allowed region.

\section{Discussion}
\label{sec:discussion}
 
In this work we computed universal bounds on scaling dimensions and central charges of 3d parity-invariant CFTs containing fermions charged under an $O(N)$ symmetry. We observed a sequence of ``kinks" that match the $O(N)$ Gross-Neveu fixed points at large $N$ and can potentially be used to learn about the theories at small $N$. At larger values of the fermion dimension, we also observed a sequence of discontinuous jumps in the bounds occurring when some scalar operator dimension passes through marginality. It will be important in future work to clarify two questions: 1) Why do these jumps always coincide with an operator becoming marginal and can this be understood analytically? 2) Do these jumps coincide with new physical ``dead-end" CFTs? One avenue for making progress on these questions is to more carefully study extremal spectra as one passes through these discontinuities.
 
We would also like to understand how to isolate the GNY (and possibly GNY*) fixed points as closed islands in the allowed space of scaling dimensions, similar to how the $O(N)$ vector models were isolated in~\cite{Kos:2015mba,Kos:2016ysd}. This will likely require extending the bootstrap to mixed correlators containing both fermions and scalars, for which the needed conformal blocks were worked out in~\cite{Iliesiu:2015akf}. An important question is whether the GNY* fixed points are fully unitary in 3d or whether they are only approximately unitary. If the GNY* kinks could be turned into closed islands then a way to probe this question would be to push the bootstrap to higher derivative order and check if the islands eventually disappear. Along these lines it will also be interesting to isolate and learn more about the 3d $\cN=1$ supersymmetric extension of the Ising model, and to find ways to probe important variants of the GNY models with multiple scalar order parameters (e.g., the variant with 3 scalars is connected to the Hubbard model on a honeycomb lattice~\cite{Sachdev:2010uz}) or additional supersymmetry.

Finally, it could be interesting to relax the assumption of parity and probe 3d fermionic CFTs with parity violation. It should also be possible to perform a similar study of the fermion bootstrap in 4d, making contact with BSM ideas related to partial compositeness~\cite{Agashe:2004rs,Caracciolo:2012je,Caracciolo:2014cxa}. We hope that our study helps to illustrate that the spinning bootstrap, even with additional global symmetry structure, is currently viable and that many more nontrivial results may be attained by studying other external spinning operators such as fermions in other global symmetry representations, global symmetry currents, stress-energy tensors, and more.
 
\section*{Acknowledgments}
We thank Simone Giombi and Igor Klebanov for discussions, and Ran Yacoby for many discussions and collaboration in the early stages of this work. Special thanks also to Grigory Tarnopolsky for adapting the three-loop $\epsilon$-expansion results of \cite{2017arXiv170308801M} for the GNY$^*$ model and for numerous other useful discussions.  LVI and SSP are supported in part by the US NSF under grant No.~PHY-1418069 and by the Simons Foundation grant No.~488653.  DP is supported by NSF grant PHY-1350180 and Simons Foundation grant 488651.  DSD is supported by DOE grant DE-SC0009988, a William D. Loughlin Membership at the Institute for Advanced Study, and Simons Foundation grant 488657 (Simons Collaboration on the Non-perturbative Bootstrap). The computations in this paper were run on the Feynman and Della clusters supported by Princeton University, the Omega and Grace computing clusters supported by the facilities and staff of the Yale University Faculty of Arts and Sciences High Performance Computing Center, Savio computational cluster resource provided by the Berkeley Research Computing program at the University of California Berkeley, as well as the Hyperion computing cluster supported by the School of Natural Sciences Computing Staff at the Institute for Advanced Study.

\appendix

\section{Implementation in \texttt{SDPB}}
\label{app:sdpb}

We now give a description of the numerical implementation of the $O(N)$-symmetric fermionic bootstrap, using \SDPB\ to search for functionals $\vec \alpha_i$ satisfying constraints such as (\ref{eq:properties}) or (\ref{eq:constraintsonfunctionalforcentralcharge})  \cite{Simmons-Duffin:2015qma}.  In order to implement a semi-definite program we limit our search over the space of functionals $\alpha_{I_\pm}$ that take the form:
\begin{equation}
\vec \alpha_{I_{\pm}}[\vec f] = \sum_{\substack{n \leq m, \\ m+n \leq \Lambda} } \vec a_{mn}^{I_\pm} \left(\partial_z^m \partial_{\overline{z}}^n \vec f(z, \overline{z})\right)
\bigg|_{z=\overline{z} = \dfrac{1}{2}} \,,
\end{equation} 
with $u = z \overline{z}$ and $v =(1-z) (1-\overline{z})$ and have evaluated the vector of functions $\vec f$ at the crossing symmetric point $z=\overline{z} = {1}/{2}$.

Applying these functionals to our crossing equations involves the $(z, \bar z)$ derivatives of the functions $g^{I_\pm, R}$ appearing in the vectors $\vec V^{I_\pm, R}$ \eqref{eq:def-of-Vs}.  As noticed in our previous work \cite{Iliesiu:2015qra}, these functions will have singularities as $z\to \bar z$, which we will avoid by multiplying the crossing equation by $(z - \bar z)^5$ before applying the functional $\vec \alpha^{I_\pm}$.
The derivatives of the conformal blocks $g^{I_\pm}$ for the fermionic four point functions  were determined using a \texttt{Mathematica} script previously used when computing fermionic bounds without imposing an $O(N)$-symmetry explicitly.  Using this script one can consequently write, at the crossing symmetric point $z = \bar z = 1/2$,
\begin{equation}
\partial_z^m \partial_{\overline{z}}^n {F}_{ab, \Delta, \ell}^{I_\pm}(z, \overline{z})
|_{z=\overline{z} = {1}/{2}}  \approx \chi_{\ell}(\Delta) P_{ab, \ell}^{(m, n), I^\pm}(\Delta) \,,
\end{equation}
where $P_{ab, \ell}^{(m, n), I^\pm}(\Delta)$ for $a, b \in \{1, 2\}$ or $(a, b) = (3, 3)$, $(a, b) =(4,4)$, are polynomials in $\Delta$ determined in \texttt{Mathematica} using the set of differential operators that relates  the fermionic blocks to the rational approximation of the scalar conformal blocks \cite{Iliesiu:2015qra}. Similarly, we can then write
\begin{equation}
\partial_z^m \partial_{\overline{z}}^n \vec{V}_{ab, \Delta, \ell}^{I_\pm, R}(z, \overline{z})
|_{z=\overline{z} = {1}/{2}}  \approx \chi_{\ell}(\Delta) \vec P_{ab, \ell}^{(m, n), I^\pm, R}(\Delta) \,,
\end{equation}
where $\vec P_{ab, \ell}^{(m, n), I^\pm, R}(\Delta)$ are now vectors of polynomials whose elements are simply related to $P_{ab, \ell}^{(m, n), I^\pm}(\Delta)$ by using Eq. (\ref{eq:def-of-Vs}).

 We can approximate the set of constrains (\ref{eq:properties}) and (\ref{eq:constraintsonfunctionalforcentralcharge}) in the form of a polynomial matrix program solvable using \SDPB\ \cite{Simmons-Duffin:2015qma}, 
\be
&&\text{Find } \vec a_{mn}^{I_{\pm}} \text{ such that:}\nn\\
&&-\sum_{a,b=1,2} \lambda_{\cO_0}^a \lambda_{\cO_0}^b Y_{ab,\ell_0}^{R_{\cO_0}}(\De_0) = 1 \,, \nn\\
&&Y_{ab, \ell}^R(\Delta) \succeq 0\,\,\, \,, \nn \\
&&Y_{33, \ell}^R(\Delta) \geq 0\,\,\, \,, \nn \\
&&Y_{44, \ell}^R(\Delta) \geq 0\,\,\, \text{ for all parity-odd operators with } \ell \text{ odd} \,, \label{eq:constraints}
\ee
where the $Y_{ab, \ell}$ are polynomials defined as
\be
Y_{ab, \ell} = \sum_{m,n,I_\pm} a_{mn}^{I_{\pm}}  P_{ab, \ell}^{(m, n), I^\pm} \,,
\ee
for $a, b \in \{1, 2\}$ or $(a, b) = (3, 3)$, $(a, b) =(4,4)$. In our applications, we take the operator $\cO_0$ used for the normalization of the functionals to be either the identity operator or the stress-energy tensor. Note that because of the multiplication of crossing equation by $(z - \bar z)^5$, some of the constraints in (\ref{eq:constraints}) are identically zero, or their linear combinations are identically zero, i.e.\ the set of constraints is not linearly independent. This can cause instabilities in \SDPB, making it run indefinitely.  We want to remove such ``flat directions" and give only linearly independent constraints to \SDPB\@. This can be done numerically. We can view the set of constraints (\ref{eq:constraints}) as a matrix with rows labeling the constraints and columns labeling the components of a functional, $a_{mn}^{I_{\pm}}$. We then only need to find the linearly independent rows of the matrix. That can be done for example in \texttt{Mathematica} using the built-in \texttt{RowReduce} function. Notice that this step needs to be done only once for a given $\Lambda$.

The full description of implementing the polynomial matrix program required to find $a_{mn}^{I_{\pm}}$ can be found in the \SDPB\ manual \cite{Simmons-Duffin:2015qma}. We have used a \texttt{Mathematica} script to manipulate the fermionic conformal blocks to obtain the matrix input for \SDPB\@. 
 
In order to obtain numerically accurate results we have used the parameters presented in Table~\ref{tab:parameters} in our \SDPB\ implementation. For $\Lambda = 19$ generating the input file required by \SDPB\ takes about 30 minutes (on a single core), while solving each semi-definite program takes about 1 hour (allowed points) or 10 hours (disallowed points) on an 16 core machine. 

 \begin{table}[!t]
\centering
\begin{tabular}{l c }
\hline\hline
$\Lambda$ & 19  \\
$\numax$ & 22\\
spins  & $S_{19}$ \\
\texttt{precision}  & 896 \\
\texttt{dualityGapThreshold} & $10^{-10}$\\
\texttt{primalErrorThreshold}  & $10^{-35}$ \\
\texttt{dualErrorThreshold} & $10^{-35}$ \\
\texttt{initialMatrixScalePrimal} ($\Omega_\cP$)  & $10^{40}$ \\
\texttt{initialMatrixScaleDual} ($\Omega_\cD$)  & $10^{40}$ \\
\texttt{feasibleCenteringParameter} ($\beta_\mathrm{feasible}$)  & 0.1 \\
\texttt{infeasibleCenteringParameter} ($\beta_\mathrm{infeasible}$) & 0.3\\
\texttt{stepLengthReduction} ($\gamma$)  & 0.7 \\
\texttt{choleskyStabilizeThreshold} ($\theta$) & $10^{-40}$  \\
\texttt{maxComplementarity}  & $10^{130}$ \\
\hline\hline
\end{tabular}
\caption{Parameters for the computations in this work.  Only \SDPB\ parameters that affect the numerics (as opposed to parameters like \texttt{maxThreads} and \texttt{maxRuntime}) are included.  The set of spins used is $S_{19}= \{0, 1, 2, \dots, 25\} \cup \{29,30,33,34,37,38,41,42,45,46,49,50\}$.
}
\label{tab:parameters}
\end{table}

\bibliography{Biblio}{}
\bibliographystyle{utphys}

\end{document}